# Title: Infrared Plasmons Propagate through a Hyperbolic Nodal Metal

**Authors:** Yinming Shao[1], Aaron J. Sternbach[1], Brian S. Y. Kim[2], Andrey A. Rikhter[3], Xinyi Xu[2], Umberto De Giovannini[4], Ran Jing[1], Sang Hoon Chae[2], Zhiyuan Sun[1], Seng Huat Lee[5,6], Yanglin Zhu[5,6], Zhiqiang Mao[5,6], J. Hone[2], Raquel Queiroz[1], A. J. Millis[1,7], P. James Schuck[2], A. Rubio[4,7], M. M. Fogler[3], D. N. Basov[1]

**Affiliations:**

[1]Department of Physics, Columbia University, New York, NY, 10027, USA

[2]Department of Mechanical Engineering, Columbia University, New York, NY, 10027, USA

[3]Department of Physics, University of California, San Diego, La Jolla, CA, 92093, USA

[4]Max Planck Institute for the Structure and Dynamics of Matter, Center for Free Electron Laser Science, Hamburg 22761, Germanys

[5]Department of Physics, Pennsylvania State University, University Park, PA, 16802, USA

[6]2D Crystal Consortium, Materials Research Institute, Pennsylvania State University, University Park, PA, USA

[7]Center for Computational Quantum Physics (CCQ), Flatiron Institute, New York, NY, 10010, USA

**Abstract:** Metals are canonical plasmonic media at infrared and optical wavelengths, allowing one to guide and manipulate light at the nano-scale. A special form of optical waveguiding is afforded by highly anisotropic crystals revealing the opposite signs of the dielectric functions along orthogonal directions. These media are classified as hyperbolic and include crystalline insulators, semiconductors and artificial metamaterials. Layered anisotropic metals are also anticipated to support hyperbolic waveguiding. Yet this behavior remains elusive, primarily because interband losses arrest the propagation of infrared modes. Here, we report on the observation of propagating hyperbolic waves in a prototypical layered nodal-line semimetal ZrSiSe. The observed waveguiding originates from polaritonic hybridization between near-infrared light and nodal-line plasmons. Unique nodal electronic structures simultaneously suppress interband loss and boost the plasmonic response, ultimately enabling the propagation of infrared modes through the bulk of the crystal.

**One-Sentence Summary:** Nodal-line metals allow hyperbolic infrared waveguiding through the bulk with band-structure-engineered loss reduction.

**Main Text:**

Nodal-line semimetals host Dirac-like linear dispersion of electronic bands with nodes extending along lines/loops in the Brillouin zone (Fig. 1A) (*1*, *2*). These systems present an appealing platform to investigate quantum effects originating from the interplay of topology, reduced dimensionality and electronic correlations encoded in unconventional optical responses (*3*, *4*). Here we focus on the nodal metal ZrSiSe, which supports nearly two-dimensional electronic structure and high-mobility Dirac fermions (*2*). We show that the nodal band structure and the attendant van Hove singularities suppress the interband transitions (*5–7*) and boost plasmonic response, thus enabling propagation of infrared waveguide modes in crystalline samples. We use scanning near-field optical microscopy (SNOM) to visualize the nano-scale infrared signatures of these modes and evaluate their energy-momentum $(\omega, q)$ dispersion.

Common materials bounce light as the real part of the dielectric function ($\varepsilon = \varepsilon_1 + i\varepsilon_2$) becomes negative. Perhaps counterintuitively, anisotropic media, including layered crystals, do support propagating modes provided in-plane and out-of-plane dielectric functions are of the opposite sign ($\varepsilon_1^{ab} \cdot \varepsilon_1^c < 0$). Because the relevant isofrequency surface (Fig. 1B) is the hyperboloid such media are referred to as hyperbolic (*8–10*). In the hyperbolic regime, the interaction of light with collective modes of crystals yields hyperbolic polaritons with exotic properties, including ray-like waveguiding in the bulk. Such waveguiding has mostly been explored in polar insulators inside narrow phonon Reststrahlen bands, including: hBN (*11*, *12*), MoO$_3$ (*13*, *14*), V$_2$O$_5$ (*15*), calcite (*16*), Ga$_2$O$_3$ (*17*) and semiconducting WSe$_2$ (*18*). Hyperbolic waveguiding is anticipated in a wide variety of anisotropic conductors as well (*8*, *19–23*). As the screened plasma frequency $\omega_p$ marks the zero-crossing of $\varepsilon_1$, a vast frequency range of hyperbolicity appears between $\omega_p^c < \omega < \omega_p^{ab}$ where $\varepsilon_1^{ab} < 0$ and $\varepsilon_1^c > 0$ (*24*). While anisotropic metals in principle offer broadband hyperbolicity, the inherently strong electronic loss (*25*) prevents waveguiding. Here, we show that nodal band structure and attendant van Hove singularities dramatically enhance the plasmonic properties of a semi-metal while simultaneously reducing interband losses to the level required to observe propagating hyperbolic plasmon polaritons (HPPs) in infrared.

To visualize infrared waveguide modes in ZrSiSe, we performed two types of nano-imaging experiments (Fig. 1C). The first one involved placing thin crystals of ZrSiSe on patterned gold antennas, which served as launchers of hyperbolic rays into the interior of the sample (*18*, *26*, *27*). The second approach utilized the sample edge to reflect the HPPs and revealed characteristic higher-order hyperbolic modes (*12*, *26*). The two complementary experiments produced consistent results.

We first focus on experiments involving a Au disk launcher underneath the crystal. As illustrated in Fig. 1C, the HPPs propagate as conical rays and emerge on the top surface of the sample as "hot rings" with enhanced nano-optical contrast surrounding the edge of the Au antennas. The propagation angle θ (with respect to surface normal) is controlled by the anisotropic permittivities of the sample (*18*, *26–28*):

$$\tan(\theta) = \sqrt{-\varepsilon_1^{ab}/\varepsilon_1^c} = \frac{\delta/2}{d}, \quad (1)$$

where δ is the separation of the rings on the top surface and *d* is the sample thickness. We obtained co-located topography and nano-optical amplitude ($S_3(\omega)$; see Materials and Methods) images of a thin ZrSiSe crystal partially covering the gold disk (Fig. 1D). At $\omega = 6667$ cm$^{-1}$, a clear double-ring pattern (Fig. 1D) emerges along the Au antenna boundary.

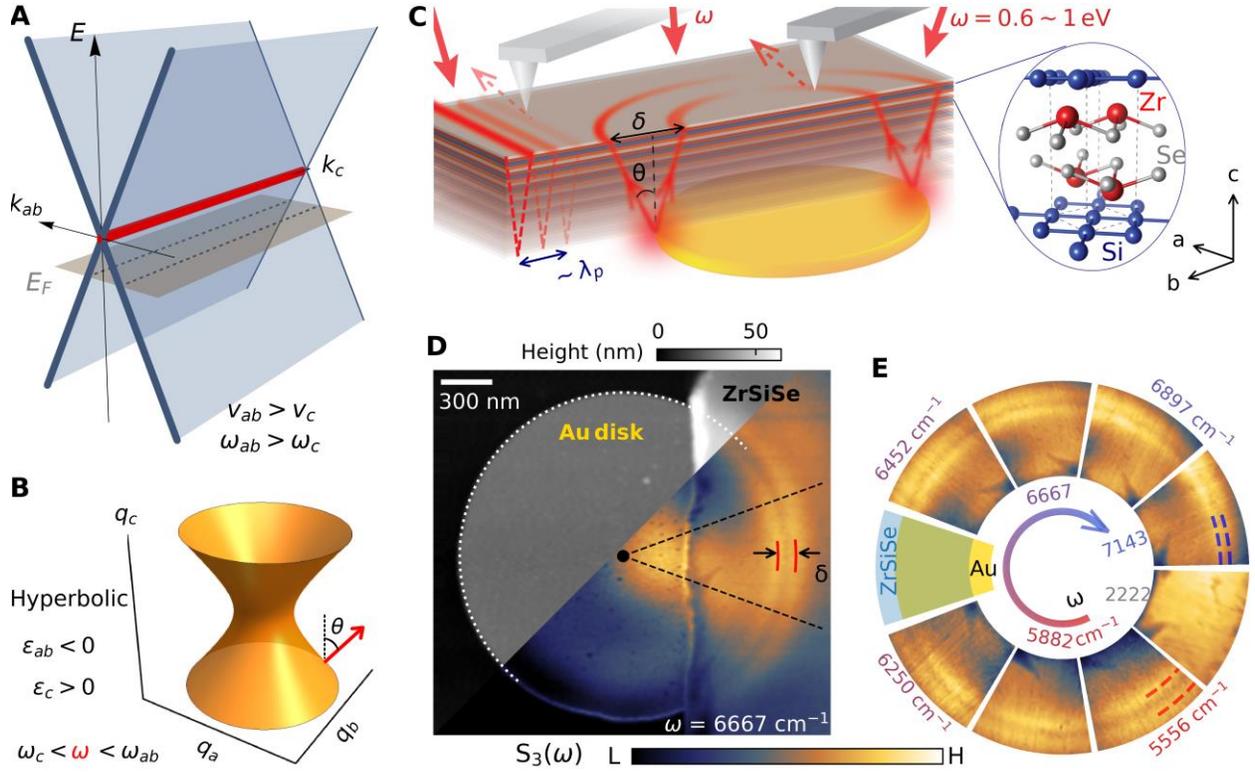

**Fig. 1. Infrared waveguide modes in nodal-line metal ZrSiSe**. **A**, Schematic band structure $E$ vs $k_{ab}$, $k_c$ showing the Dirac nodal-line (red). The gray plane indicates the Fermi level. **B**, Schematic isofrequency surface inside the hyperbolic regime ($\varepsilon_{ab} \cdot \varepsilon_c < 0$) for a nodal-line metal. Red arrow indicates the direction of the group velocity of the hyperbolic ray. **C**, Schematic of the nano-imaging setup. The near-infrared laser illuminates the sample and hyperbolic plasmon polaritons (red lines) are launched by an atomic-force microscope (AFM) tip at the edge or by an underlying gold antenna. The AFM-based nano-optics registers the evanescent fields associated with the waveguide modes in the bulk in the form of linear fringes or characteristic rings. The layered crystal structure of ZrSiSe is shown in the inset. **D**, Topography (gray-scale) and near-field scattering amplitude $S_3$ (color-scale) of a 26 nm ZrSiSe crystal partially covering a gold disk. White dotted line indicates the boundary of the Au disk. Red solid lines mark the split center-peaks of hyperbolic polariton modes along the circumference. **E**, Images of $S_3$ obtained within the sector region indicated by black dashed lines in (D) and assembled for laser frequencies from $\omega = 7143$ cm$^{-1}$ to 5556 cm$^{-1}$ within the hyperbolic region. The image taken outside of the hyperbolic range at 2222 cm$^{-1}$ is devoid of the double-ring structure.

This double-ring pattern is confined to the vicinity of the antenna edges and is distinct from the intensity variation in the interior of our structures prompted by the internal resonances of the Au antenna at much longer length scales. In Fig. 1D, the ring separation ($\delta \approx 150$ nm) is an order of magnitude smaller than the free-space light wavelength ($\lambda = 1.5$ μm, $\omega \approx 6667$ cm$^{-1}$). The double-ring pattern also varies with incident light frequency, as shown in Fig. 1E where we assemble the $S_3(\omega)$ data at selected frequencies. The blue and red dashed lines mark the positions of the hot rings at $\omega = 7143$ cm$^{-1}$ and $\omega = 5556$ cm$^{-1}$, with systematic evolution of the ring separation for frequencies in between. In contrast, the double-ring feature is completely absent in the sector for $\omega = 2222$ cm$^{-1}$ outside of the hyperbolic range quantified in Fig. 2; instead, this sector shows a homogeneous near-field response (see also Fig. S6).

We now inquire into quantitative details of propagating HPPs in ZrSiSe. We average the radial line profiles within the sectors depicted in Fig. 1E and plot these in Fig. 2A. The experimental ring separation δ(ω) is obtained by fitting the line profile with two Gaussian functions and a linear background, shown for ω = 7634 cm$^{-1}$ and 5556 cm$^{-1}$ in Fig. 2A (see supplementary text Sec. S3 for complete analysis). With the *ab*-plane permittivity known from Ref. (*5*), the experimental ring-separation δ together with the sample thickness *d* allows for the extraction of the *c*-axis permittivity of ZrSiSe from Eq. (1). These latter data are displayed in Fig. 2B along with the experimental *ab*-plane permittivity (squares). The hyperbolic regime in ZrSiSe extends between ≈ 2837 − 9091 cm$^{-1}$ (see supplementary text Sec. S1, S3). Finally, we observe yet another hallmark of the hyperbolic rays, which is the scaling of the inter-peak separation δ with increasing sample thickness (Fig. 2C), $\delta = 2d\sqrt{-\varepsilon_1^{ab}/\varepsilon_1^c}$. Broadband hyperbolic electrodynamics in the layered nodal-metal ZrSiSe is therefore firmly established.

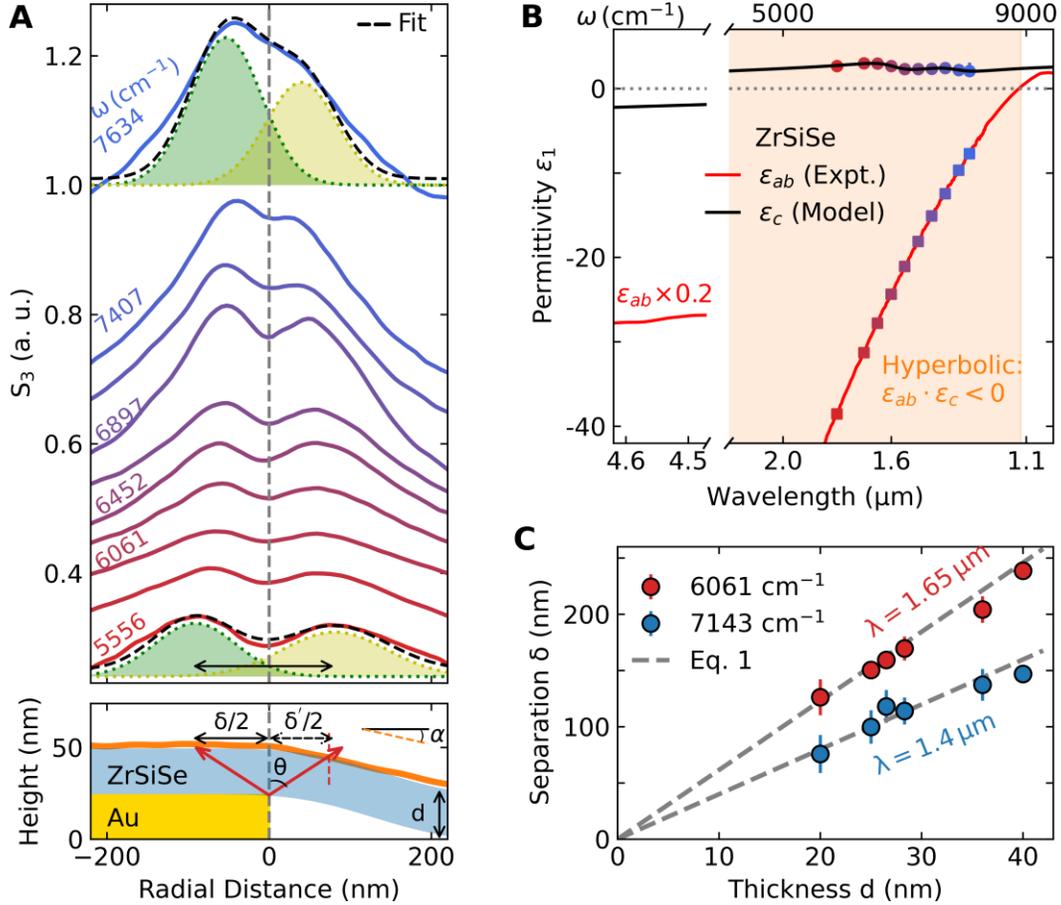

**Figure 2| Hyperbolic electrodynamics of ZrSiSe**. (**A**), Line profiles of the near-field scattering amplitude $S_3$ at several incident frequencies. Black dashed lines are fits using Gaussian functions for the ω = 7634 cm$^{-1}$ and 5556 cm$^{-1}$ line profiles. Green and yellow shaded areas indicate the individual Gaussian functions representing the hyperbolic ray profiles. Bottom panel shows the topography line profile (orange) near the edge of the Au disk (gold) in Fig.1D. The downward slope (tan α) of the sample (blue) leads to a geometrical correction to the measured ring-separation δ. (**B**), In-plane dielectric function ($\varepsilon_{ab}$, red line) obtained from far-field optical measurements (*5*). The $\varepsilon_{ab}$ values at selected frequencies (squares) together with δ(ω) in (A) are used to extract $\varepsilon_c$ (circles) using Eq. 1. The black line is a Drude-Lorentz fit of the experimental out-of-plane dielectric function data. (**C**), Hyperbolic ray separation δ(ω) as a function of flake thickness *d* at ω = 6061 cm$^{-1}$ (red) and ω = 7143 cm$^{-1}$ (blue). The separations scale linearly with increasing flake thickness, as prescribed by Eq. 1 (grey dashed line).

The natural edges of thin hyperbolic materials can also launch and reflect polaritons emanating from the metallic tip (Fig. 1C) (29). To explore HPPs near the edges, we focused on the phase contrast, which provides highest level of image fidelity. The phase-contrast data reveal weak higher-order HPP modes: yet another electrodynamics signature of hyperbolicity (12, 26). In Figures 3A-3C, we present the topography and near-field phase-contrast images obtained for a 20 nm thin ZrSiSe crystal on a Si/SiO$_2$ substrate for two representative laser frequencies. At $\omega = 8333$ cm$^{-1}$ (Fig. 3B), the phase-contrast displays a prominent fringe near the edge, which shifts further into the interior of the sample as the laser frequency decreases in Fig. 3C. The first peak-dip separation systematically increases in thicker samples (Fig. S15). In order to quantify the HPP wavelength, we utilized a previously developed electromagnetic solver (30) to simulate the phase contrast with the complex polariton momentum $q_p = (1 + i\gamma)2\pi/\lambda_p$ as input. Here, $\lambda_p$ is the polariton wavelength and $\gamma$ accounts for the damping of the polariton wave.

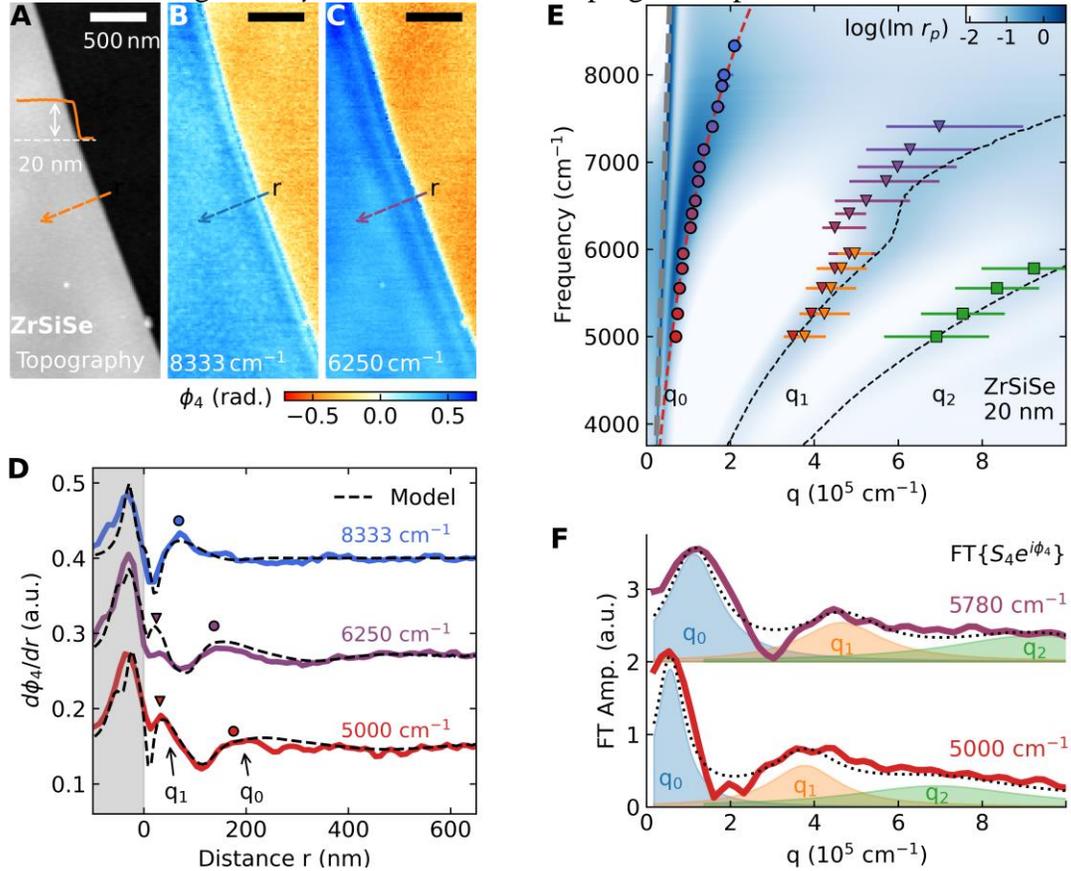

**Fig. 3. Hyperbolic plasmon polaritons in ZrSiSe**. Topography (**A**) and near-field phase ($\phi_4$) image of a 20 nm thin crystal of ZrSiSe at (**B**) $\omega = 8333$ cm$^{-1}$ and (**C**) $\omega = 6250$ cm$^{-1}$. (**D**), Phase derivative line profiles ($d\phi_4/dr$) at multiple laser frequencies near edges of ZrSiSe. For $\omega = 6250$ cm$^{-1}$ and 5000 cm$^{-1}$, the derivative profiles reveal features at multiple spatial periodicities ($q_0$ and $q_1$). Black dashed lines are the simulation of the derivative profile with two periodicities, corresponding to the principal ($q_0$) and higher-order ($q_1$) HPPs. (**E**), Frequency-momentum dispersion of HPP plotted in the form of Im(r$_p$). Circles: the principal modes; triangles: higher-order polaritons. Data points are superimposed over the calculated Im(r$_p$) described in the text. The grey dashed line represents the free-space light cone. Black dashed lines indicate numerical solutions for the divergence of Im(r$_p$) for the higher-order HPP branches. Red dashed line is a guide for the dispersion of the principal branch. (**F**), Fourier transform (FT) of the complex near-field signal $S_4 e^{i\phi_4}$ along the same path in (B-D) at $\omega = 5780$ cm$^{-1}$ and 5000 cm$^{-1}$. Multiple peaks in the FT amplitude correspond to the principal ($q_0$) and higher-order ($q_1$, $q_2$) modes and are fitted by Lorentzian functions (color-shaded area).

Multiple fringes of different periodicity appear at lower frequencies and are particularly apparent at $\omega = 6250$ cm$^{-1}$ (Fig. 3C). To better resolve these shorter wavelength oscillations, we inspected the derivative of the phase line profiles, $d\phi_4/dr$. In Fig. 3D, we show the experimental and simulated phase derivative traces for $\omega = 8333$ cm$^{-1}$, 6250 cm$^{-1}$ and 5000 cm$^{-1}$ (see Fig. S13 for additional data). The profile obtained at $\omega = 8333$ cm$^{-1}$ can be adequately reproduced with a single damped polariton of wavelength $\lambda_{p0} \approx 300\ nm$. However, for $\omega = 6250$ cm$^{-1}$ and 5000 cm$^{-1}$, an additional mode with a much shorter wavelength $\lambda_{p1}$ is needed to fully account for the data. Both weaker ($q_1$) and stronger ($q_0$) peaks in the derivative line profiles are well described by the simulation (black dashed line) involving two polariton modes with different polaritons of wavelengths $\lambda_{p0}$ and $\lambda_{p1}$. The polariton momentum can then be extracted from the wavelength as Re $q_p = \frac{2\pi}{\lambda_p}$, enabling a direct comparison with theoretical dispersion. As we show below, the two modes correspond to the principal and higher-order HPPs in ZrSiSe and are in accord with the experimental dielectric tensors in Fig. 2B.

The extracted HPP momenta are organized in the dispersion $(\omega, q)$ plot in Fig. 3E. It is customary to identify HPPs via the divergences of the reflection coefficient $r_p(\omega, q)$ (*11*). A colormap of Im($r_p$) provides an instructive way to visualize both the dispersion and the damping of the HPP modes. The colormap is calculated for a 20 nm thick crystal of ZrSiSe residing on a SiO$_2$/Si substrate using experimental dielectric functions (Fig. 2B). As expected, multiple dispersive branches develop in the hyperbolic frequency range, corresponding to the principal and higher-order modes. The existence of higher-order modes can also be documented by directly Fourier transforming the experimental real-space line profile (*12*, *26*). As shown in Fig. 3F, the Fourier transform amplitudes of the complex signal $S_4 e^{i\phi_4}$ for $\omega = 5780$ cm$^{-1}$ and 5000 cm$^{-1}$ (see Fig. S14 for additional data) indeed display up to three distinct modes that can be parameterized by Lorentzian functions. The obtained momenta $(q_0, q_1)$ are consistent with the values from line profile modeling (Fig. 3D, 3E); Fourier transforms are also suggestive of an additional weaker higher-order mode $q_2$. The calculated hyperbolic dispersions agree with the experimental momenta (colored circles and triangles), unequivocally corroborating the notion of hyperbolic plasmon polaritons in ZrSiSe. The deviation of the higher-order branch ($q_1$) and the data points (triangles) is within the experimental error bars. Nevertheless, this slight discrepancy hints at the presence of surface states in ZrSiSe (*31–33*) with potentially different dielectric responses (see supplementary text S4) from the bulk values (Fig. 2B) used in our calculation.

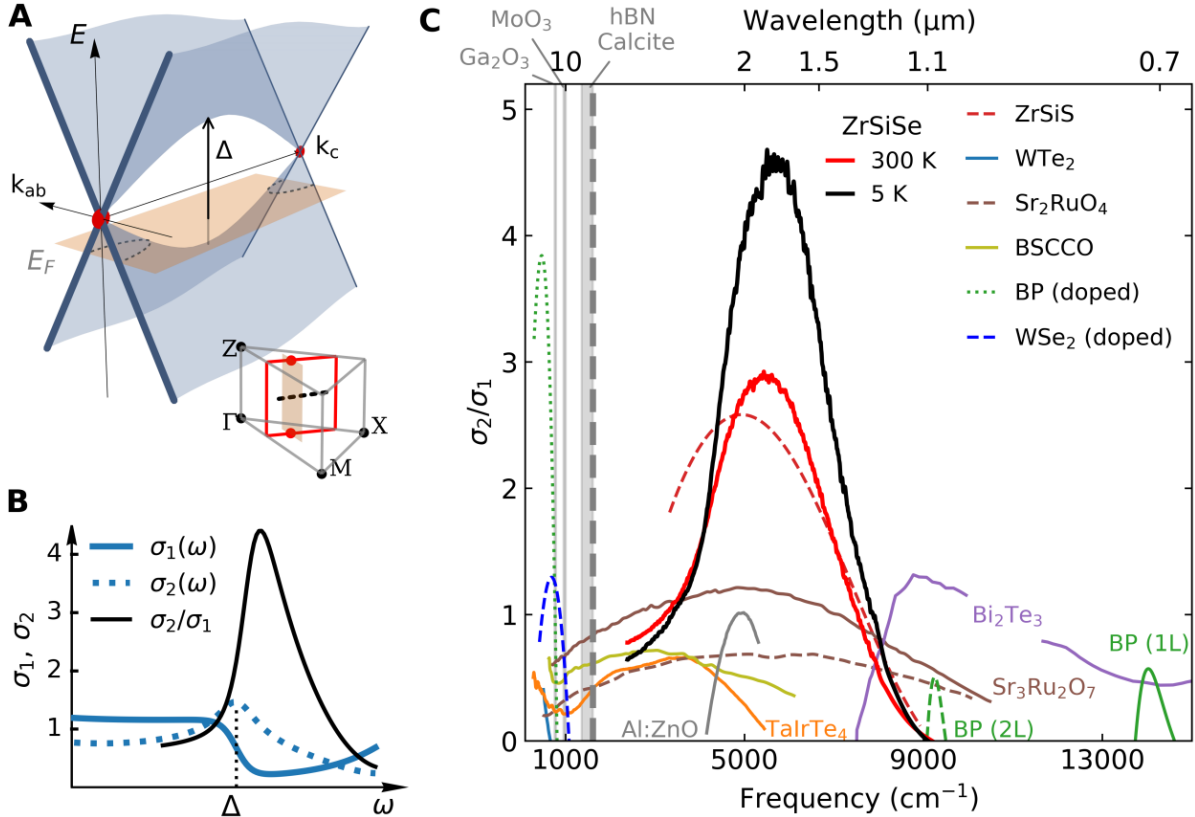

**Fig. 4. Band structure origin of enhanced plasmonic hyperbolicity in a nodal-metal**. (**A**), Schematic band structure E vs $k_{ab}$, $k_c$ inside the nodal-square (red line in inset). Vertical arrow indicates the van Hove energy Δ. (**B**), Model for the interband optical conductivity ($\sigma = \sigma_1 + i\sigma_2$) of a nodal-line near the van Hove energy (see Materials and Methods). The ratio $\sigma_2/\sigma_1$ exhibits a maximum near Δ. The optical conductivity is related to the dielectric function via $\sigma = i\omega(1-\varepsilon)/4\pi$. (**C**), Survey of experimental $\sigma_2/\sigma_1$ ratio for representative plasmonic hyperbolic materials (WTe$_2$ (*19*, *20*), TaIrTe$_4$ (*21*), ZrSiS (*22*), black phosphorus (BP) (*23*), doped WSe$_2$ (*18*), Bi$_2$Sr$_2$CaCu$_2$O$_8$ (*8*, *34*), Sr$_2$RuO$_4$ (*35*)) and excitonic hyperbolic materials (few-layer BP (*36*)). (See Fig. S20-S27 for extracted optical conductivities)

The propagating HPPs observed in ZrSiSe would not have been possible without tamed interband losses: a unique feat of the nodal band structure uncovered by our experiments. In ZrSiSe, the nodal-lines form "squares" in momentum space with lines of van Hove singularities inside the nodal-squares (Fig. 4A inset) (*5*, *7*). The resulting saddle-point structure (Fig. 4A) leads to a suppression of interband transitions above the van Hove energy (Δ). The corresponding dissipative part of the conductivity $\sigma_1(\omega)$ shows a "cliff" above Δ (Fig. 4B), accompanied by a peak in $\sigma_2(\omega)$ at Δ prescribed by Kramers-Kronig relations. We emphasize that the enhancement in $\sigma_2(\omega)$ is an important approach towards high-quality plasmons (*25*, *30*, *37*). The unique combination of reduced dissipation ($\sigma_1$) and enhanced plasmonic response ($\sigma_2$), quantified by the ratio $\sigma_2/\sigma_1 \approx 3-5$ for ZrSiSe (Fig. 4C) is superior to that of all other candidate plasmonic and excitonic hyperbolic materials reported so far.

The suppression of interband transitions near the van Hove energy offers a concrete strategy for the "band structure engineering" approach to mitigating loss and boosting plasmonic response (*38*, *39*). While the existence of "lossless metal" remains elusive (*25*), we propose that van Hove singularities in topological systems (*40*) reveal as-yet untapped plasmonic design rule afforded by nodal-line semimetals.

**References and Notes**

1. L. M. Schoop *et al.*, *Nat. Commun.* **7**, 11696 (2016).

2. J. Hu *et al.*, *Phys. Rev. Lett.* **117**, 016602 (2016).

3. N. P. Armitage, E. J. Mele, A. Vishwanath, *Rev. Mod. Phys.* **90**, 015001 (2018).

4. J. Orenstein *et al.*, *Annu. Rev. Condens. Matter Phys.* **12**, 247–272 (2021).

5. Y. Shao *et al.*, *Nat. Phys.* **16**, 636–641 (2020).

6. M. B. Schilling, L. M. Schoop, B. V. Lotsch, M. Dressel, A. V. Pronin, *Phys. Rev. Lett.* **119**, 187401 (2017).

7. T. Habe, M. Koshino, *Phys. Rev. B*. **98**, 125201 (2018).

8. E. E. Narimanov, A. V. Kildishev, *Nat. Photon.* **9**, 214–216 (2015).

9. H. N. S. Krishnamoorthy, Z. Jacob, E. Narimanov, I. Kretzschmar, V. M. Menon, *Science*. **336**, 205–209 (2012).

10. A. V. Kildishev, A. Boltasseva, V. M. Shalaev, *Science*. **339**, 1232009 (2013).

11. S. Dai *et al.*, *Science*. **343**, 1125–1129 (2014).

12. A. J. Giles *et al.*, *Nat. Mater.* **17**, 134–139 (2018).

13. W. Ma *et al.*, *Nature*. **562**, 557 (2018).

14. Z. Zheng *et al.*, *Sci. Adv.* **5**, eaav8690 (2019).

15. J. Taboada-Gutiérrez *et al.*, *Nat. Mater.* **19**, 964–968 (2020).

16. W. Ma *et al.*, *Nature*. **596**, 362–366 (2021).

17. N. C. Passler *et al.*, *Nature*. **602**, 595–600 (2022).

18. A. J. Sternbach *et al.*, *Science*. **371**, 617–620 (2021).

19. A. J. Frenzel *et al.*, *Phys. Rev. B*. **95**, 245140 (2017).

20. C. Wang *et al.*, *Nat. Comm.* **11**, 1158 (2020).

21. Y. Shao *et al.*, *Proc. Natl. Acad. Sci. U. S. A.* **118**, e2116366118 (2021).

22. J. Ebad-Allah *et al.*, *Phys. Rev. Lett.* **127**, 076402 (2021).

23. S. Biswas *et al.*, *Sci. Adv.* **7**, eabd4623 (2021).

24. T. Low *et al.*, *Nat. Mater.* **16**, 182–194 (2017).

25. A. Boltasseva, H. A. Atwater, *Science*. **331**, 290–291 (2011).

26. S. Dai *et al.*, *Nat. Commun.* **6**, 6963 (2015).

27. P. Li *et al.*, *Nat. Commun.* **6**, 7507 (2015).

28. J. D. Caldwell *et al.*, *Nat. Commun.* **5**, ncomms6221 (2014).

29. S. Dai *et al.*, *Nano Lett.* **17**, 5285–5290 (2017).

30. Z. Fei *et al.*, *Nature*. **487**, 82–85 (2012).

31. Z. Zhu *et al.*, *Nat. Commun.* **9**, 4153 (2018).

32. S. Xue *et al.*, *Phys. Rev. Lett.* **127**, 186802 (2021).

33. A. Topp *et al.*, *Phys. Rev. X.* **7**, 041073 (2017).

34. M. E. Berkowitz *et al.*, *Nano Lett.* **21**, 308–316 (2021).

35. T. Katsufuji, M. Kasai, Y. Tokura, *Phys. Rev. Lett.* **76**, 126–129 (1996).

36. F. Wang *et al.*, *Nat. Comm.* **12**, 5628 (2021).

37. J. Chen *et al.*, *Nature*. **487**, 77–81 (2012).

38. M. N. Gjerding, M. Pandey, K. S. Thygesen, *Nat. Commun.* **8**, 15133 (2017).

39. C. Lewandowski, L. Levitov, *PNAS*. **116**, 20869–20874 (2019).

40. N. F. Q. Yuan, H. Isobe, L. Fu, *Nat. Comm.* **10**, 5769 (2019).

**Acknowledgments:**
**Funding:** Research in polaritons physics at Columbia is supported by DOE-BES grant DE-SC0018426.The development of nano-imaging capabilities in near-infrared is funded by the Vannevar Bush Faculty Fellowship ONR-VB: N00014-19-1-2630. D.N.B. is a Moore Investigator in Quantum Materials EPIQS GBMF9455. Non-linear imaging and spectroscopy are funded supported as part of Programmable Quantum Materials, an Energy Frontier Research Center funded by the U.S. Department of Energy (DOE), Office of Science, Basic Energy Sciences (BES), under award DE-SC0019443. Support for crystal growth and characterization at Penn State was provided by the National Science Foundation through the Penn State 2D Crystal Consortium-Materials Innovation Platform (2DCC-MIP) under NSF Cooperative Agreement DMR-1539916 and NSF-DMR 1917579. U.D.G. and A.R. acknowledge support from the European Research Council (ERC-2015-AdG-694097), Grupos Consolidados (IT1249-19), and SFB925. We acknowledge funding by the Deutsche Forschungsgemeinschaft (DFG, German Research Foundation) under Germany's Excellence Strategy - Cluster of Excellence and Advanced Imaging of Matter (AIM) EXC 2056 - 390715994 and RTG 2247. We also acknowledge the computational resource provided by the Max Planck Computing and Data Facility. This work was supported by the Max Planck-New York City Center for Nonequilibrium Quantum Phenomena. The Flatiron Institute is a division of the Simons Foundation.

**Author contributions:** Y.S. and D.N.B. conceived the study. Y.S. and A.J.S. performed the near-field experiments. B.S.Y.K fabricated the device with assistance from S.H.C. and supervised by J.H. A.A.R. performed the electrodynamic simulation with supervision from M.F.F. Y.Z and S.H.L grew the single crystals and performed transport measurements supervised by Z.M. U.D.G. performed first principal calculations with supervision from A.R. Y.S. and D.N.B. wrote the manuscript with input from all coauthors.

**Competing interests:** The authors declare no competing financial interests.

**Data and materials availability:** All data are available in the main text or the supplementary materials.

**Supplementary Materials**

Materials and Methods

Supplementary Text

Figs. S1 to S27

Tables S1

References (*1–21*)

Supplementary Materials for

# Infrared Plasmons Propagate through a Hyperbolic Nodal Metal

Yinming Shao[1], Aaron J. Sternbach[1], Brian S. Y. Kim[2], Andrey A. Rikhter[3], Xinyi Xu[2], Umberto De Giovannini[4], Ran Jing[1], Sang Hoon Chae[2], Zhiyuan Sun[1], Seng Huat Lee[5,6], Yanglin Zhu[5,6], Zhiqiang Mao[5,6], J. Hone[2], Raquel Queiroz[1], A. J. Millis[1,7], P. James Schuck[2], A. Rubio[4,7], M. M. Fogler[3], D. N. Basov[1]

[1]Department of Physics, Columbia University, New York, NY, 10027, USA

[2]Department of Mechanical Engineering, Columbia University, New York, NY, 10027, USA

[3]Department of Physics, University of California, San Diego, La Jolla, CA, 92093, USA

[4]Max Planck Institute for the Structure and Dynamics of Matter, Center for Free Electron Laser Science, Hamburg 22761, Germany

[5]Department of Physics, Pennsylvania State University, University Park, PA, 16802, USA

[6]2D Crystal Consortium, Materials Research Institute, Pennsylvania State University, University Park, PA, USA

[7]Center for Computational Quantum Physics (CCQ), Flatiron Institute, New York, NY, 10010, USA

**Materials and Methods**
**Supplementary Text**

**References**

**Materials and Methods**

Single crystal growth and device fabrication.

The ZrSiSe single crystals were synthesized using a chemical vapor transport method as described previously (*1*, *2*). For Au antenna patterned devices, Au/Cr (25 nm/1 nm) disks were e-beam deposited on $SiO_2$/Si substrates following standard e-beam lithography processes using a lift-off resist. ZrSiSe flakes were then directly exfoliated on Au/Cr disks in a glovebox filled with inert gas ($O_2$ < 1 ppm, $H_2O$ < 0.1 ppm). Prior to exfoliation, the substrates were annealed in glovebox at 250 °C for 1 hr to remove any residual moisture on the surface.

Near-infrared nano-optical measurements.

We used a scattering-type scanning near-field optical microscope (s-SNOM, Neaspec) based on an atomic force microscope (AFM) operating in tapping mode. The tapping frequency of the AFM tip is around 70 kHz and near-field data are collected at higher harmonic (n = 3 or 4) of the tapping frequency to suppress the far-field background. For the gold antenna launcher experiment the difference frequency generation outputs of a pulsed laser source (Pharos, Light Conversion) were used. For the edge-launching experiments, we used a continuous-wave tunable laser from M Squared. Tunable outputs between 1140 nm – 2200 nm are generated by frequency mixing of a high-power 532 nm diode laser (EQUINOX) and a Ti:Sapphire laser tunable between 700 nm – 1000 nm (SolsTiS).

Geometrical correction considering finite antenna thickness.

Due to the finite thickness of the underlying Au antenna, the sample exhibits a downward slope outside the antenna boundary (bottom of Fig.2A in the main text). This small slope ($\tan\alpha$) leads to a finite asymmetry ($\frac{\delta}{2} > \frac{\delta'}{2}$) in the propagation distance of the two rays (Fig. 2A), which we corrected in the extraction of δ(ω) as following. The two distances δ and δ' are related by $\frac{\delta}{\delta'} = 1 + \tan\alpha \cdot \tan\theta$, where $\tan\theta = \frac{\delta}{2d}$ and $d$ is the sample thickness. The measured double-ring distance is denoted as $\Delta = \frac{\delta}{2} + \frac{\delta'}{2}$. Substituting δ' in terms of Δ and δ, we obtained $\delta = \Delta - \frac{2d}{\tan\alpha} + \sqrt{\Delta^2 + \frac{4d^2}{\tan^2\alpha}}$. At the limit of $\alpha \to 0$ (no downward slope) or $d \to \infty$ (infinitely thick sample), $\sqrt{\Delta^2 + \frac{4d^2}{\tan^2\alpha}} \approx \frac{2d}{\tan\alpha}$ and therefore $\delta \to \Delta$ as expected.

Electronic structure calculations of ZrSiSe

The electronic structure of the system was investigated with density functional theory (DFT). DFT calculations were carried out at the level of DFT plus onsite Hubbard U and intersite V (DFT+U+V) (*3*), as implemented in the Octopus code (*4*), which delivers an hybrid-like quality of the band structure at a fraction of the computational cost (*5*). Experimental lattice constants of $a = 3.623$ Å and $c = 8.365$ Å were employed. For the slab configuration, containing 5 layers of ZrSiSe, a 16 Å vacuum region was chosen to properly converge the bands along the non-periodic dimension z. The ground state was calculated by discretizing the equations in real-space with a spacing of 0.159 Å and spin-orbit coupling was fully accounted for valence electrons while core electrons were treated with relativistic HGH pseudopotentials (*6*). The Brillouin zone was sampled with a 16×16×8 Monkhorst-Pack grid for the bulk and a 15×15 grid for the slab geometries.

Model for interband optical conductivity of ZrSiSe near van Hove singularity.

The minimum in the interband optical conductivity of ZrSiSe (see Fig. S1) can be modeled by a step-function near the van Hove singularities and a Lorentzian function accounting for transitions at higher energy. Specifically, the step-function is expressed as $\sigma_{step}(\omega) = \frac{\tanh[(-\omega+\Delta)/\Gamma]+1}{2} + \sigma_s$, where $\Delta$ is the van Hove energy, $\Gamma$ is the step-width and $\sigma_s$ is a constant background. The higher energy optical transition is described by $\sigma_{high}(\omega) = -i\omega(\varepsilon_\infty - 1 + f^2/(\omega_0^2 - \omega^2 - i\gamma\omega))$, where $\varepsilon_\infty$ is the high frequency dielectric constant, $f^2$, $\omega_0$ and $\gamma$ are the oscillator strength, center frequency and scattering rate of the Lorentzian peak, respectively. The real part of the interband optical conductivity of ZrSiSe can then be expressed as $\sigma_1 = \sigma_{step}(\omega) + Re[\sigma_{high}(\omega)]$ and the imaginary part $\sigma_2$ are obtained numerically through Kramers-Kronig relations. For modeled $\sigma_1$ and $\sigma_2$ shown in Fig. 4B of the main text, the following parameters are used: $\Delta = 1.25, \Gamma = 0.06, \sigma_s = 0.1, \omega_0 = 2, f^2 = 0.49, \gamma = 0.5$ and $\varepsilon_\infty = 2.5$.

**Supplementary Text**

Section S1: *c*-axis dielectric function and optical conductivity of ZrSiS and ZrSiSe

The *ab*-plane complex optical conductivity ($\sigma(\omega) = \sigma_1 + i\sigma_2$) of bulk ZrSiSe is obtained using broadband reflectance spectra combined with spectroscopic ellipsometry (*2*). To study the hyperbolicity the *c*-axis dielectric response is also important. While an optically flat *ac*-surface is not attainable in ZrSiSe, measurements on the large and flat *ac*-surface of a closely related ZrSiS compound indeed reveal a much lower plasma frequency along the *c*-axis (Fig. S1, black dashed line), consistent with recent report (*7*). In order to obtain the *c*-axis dielectric function of ZrSiSe, we utilized the gold antenna launchers and performed near-field imaging experiments as discussed in the main text. Given the knowledge of *ab*-plane dielectric function, the antenna launching experiment allow us to extract the *c*-axis dielectric function based on the double-ring spacing (see Section S3). The extracted *c*-axis screened plasma frequency of ZrSiSe is around 3000 $cm^{-1}$ (Fig. S5), similar to the screened plasma frequency of ZrSiS (Fig. S1).

As mentioned in the main text, the unique nodal-square structure of ZrSiSe offers an effective approach to reducing the electronic loss associated with interband optical transitions. We remark that although van Hove singularities appear in many electronic systems (e.g., Weyl semimetals), the impact on the optical conductivity of 3D systems rarely leads to a minimum, due to the large joint density of states (JDOS). For a pair of Weyl nodes, the JDOS scales with frequency as $\omega^2$ and correspondingly, $\sigma_1(\omega) \propto \frac{\text{JDOS}}{\omega} = \omega$. As a result, the van Hove singularity only changes the slope of the linear scaling of $\sigma_1$ in a Weyl semimetal (8, 9), in contrast to the minimum observed in ZrSiSe (2, 10, 11). This van Hove singularity induced suppression of interband absorption is effective for both the *ab*-plane and *c*-axis optical conductivity in ZrSiS/Se (2, 7).

Section S2: Transport measurements and carrier density of ZrSiSe

Magnetoresistivity and Hall resistivity of ZrSiSe were performed using a standard four-probe technique in a physical properties measurement system (PPMS, Quantum Design), as shown in Fig. S2. Given the coexistence of both electron and hole-like carriers in ZrSiSe, we adopt a two-band model to estimate the carrier densities and mobilities by simultaneously fitting the measured magnetoresistivity and Hall resistivity data. If the contributions of both electron and hole bands to conductivity are assumed to be additive, the longitudinal resistivity ($\rho_{xx}$) and transverse resistivity ($\rho_{xy}$) can be described by:

$$\rho_{xx} = \frac{(n_e \mu_e + n_h \mu_h) + (n_e \mu_e \mu_h^2 + n_h \mu_h \mu_e^2) B^2}{(n_e \mu_e + n_h \mu_h)^2 + \mu_e^2 \mu_h^2 (n_h - n_e)^2 B^2} \cdot \frac{1}{e} \quad (S1)$$

$$\rho_{xy} = \frac{(n_h \mu_h^2 - n_e \mu_e^2) + \mu_e^2 \mu_h^2 (n_h - n_e) B^2}{(n_e \mu_e + n_h \mu_h)^2 + \mu_e^2 \mu_h^2 (n_h - n_e)^2 B^2} \cdot \frac{B}{e} \quad (S2)$$

where $n_e$ ($n_h$) and $\mu_e$ ($\mu_h$) are the density and mobility of these electron and hole bands shown in equation (S1) and (S2), respectively. $B = \mu_o H$ and $e$ are the magnetic field strength and elementary charge. In Fig. S2, we present the two-band model fit to the $\rho_{xx}$ and $\rho_{xy}$ at $T = 100$ K at a low-field range since no satisfactory fit can be obtained for the low temperature data. We also note the $\rho_{xx}$ and $\rho_{xy}$ data in the high-field range cannot be fitted with the two-band model, probably due to the quantum effects and high order effect as described in (*1*). The best fits yield the carrier densities of $n_e \sim 4.1 \times 10^{20}$ cm$^{-3}$ and $n_h \sim 1.2 \times 10^{20}$ cm$^{-3}$ and the mobilities of $\mu_e \sim$ 1030 cm$^2$/Vs and $\mu_h \sim 4522$ cm$^2$/Vs, which is consistent with previous reports (*1*).

Section S3: Antenna launching near-field imaging data and fitting

In Fig. S3 we show the full frequency dependence of the antenna launching experimental data in the hyperbolic regime. Since the diameter of the Au disk antenna (2 µm) is comparable to the laser wavelength (1.3 − 1.8 µm), the near-field signal exhibits diffraction patterns inside the Au antenna, as shown in the main text. On the other hand, the ZrSiSe region covering the Au antenna shows an enhancement in near-field amplitude and a gradual increase in the "double-ring" separation. This separation reaches a maximum at $\omega = 5556$ cm$^{-1}$ and the length scale ($\approx 190$ nm) is an order of magnitude smaller than the laser wavelength ($\lambda = 1.8$ μm). To quantify the double-ring separations, we fitted the line profiles of $S_3$ with two Gaussian functions and a linear background, as shown in Fig. S4. Together with the slope correction discussed in the Materials and Methods section, we extracted the frequency-dependent peak separation $\delta(\omega)$, as shown in Fig. S5. The out-of-plane (*c*-axis) dielectric constant of ZrSiSe is then obtained using the experimental *ab*-plane dielectric constant and $\delta(\omega)$, according to $\sqrt{\varepsilon_c} = \sqrt{-\varepsilon_{ab}}\left(\frac{2d}{\delta}\right)$, where $d$ is the thickness of the ZrSiSe crystal.

The extracted *c*-axis dielectric functions of ZrSiSe are modeled with Drude-Lorentzian oscillators accounting for both the intraband and the interband contributions: $\varepsilon_c(\omega) = \varepsilon_\infty + \sum_j \frac{\omega_{p,j}^2}{\omega_{0,j}^2 - \omega^2 - i\gamma_j \omega}$. Here $\varepsilon_\infty$ is the high-frequency dielectric constant, $\omega_{0,j}$, $\omega_{p,j}^2$ and $\gamma_j$ are the center frequency, oscillator strength, and scattering rate of the j-th oscillator, respectively. The model (red line in Fig. S5) agrees well with the experimental data and the fitting parameters are listed in Table. S1.

In Fig. S6 – Fig. S11, we show the gold disk antenna launching experiment with ZrSiSe crystals of varying thicknesses on top.

Section S4: Hyperbolic plasmon polariton dispersion with surface states.

The momentum of the hyperbolic plasmon polariton (HPP) modes with surface conductivity $\sigma_{2D}$ obeys the following Fabry-Perot quantization condition:

$$q_l(\omega) = \frac{i\sqrt{\varepsilon_c}}{d\sqrt{\varepsilon_{ab}}}\left[\pi l + \arctan\left[i\frac{\varepsilon_0}{\varepsilon_1}\left(1 - \frac{2q}{q_{2D}}\right)\right] + \arctan\left[i\frac{\varepsilon_2}{\varepsilon_1}\left(1 - \frac{2q}{q_{2D}}\right)\right]\right] \quad (S3)$$

where $l = 0, 1, 2$ is the mode index, $d$ is the sample thickness, $\varepsilon_0$ ($\varepsilon_2$) is the dielectric function of the top (bottom) medium, and $\varepsilon_1 = \sqrt{\varepsilon_{ab}}\sqrt{\varepsilon_c}$ is the mean dielectric function of the hyperbolic material with the in-plane ($\varepsilon_{ab} < 0$) and out-of-plane ($\varepsilon_c > 0$) components having opposite signs. Here, $q_{2D}$ is related to the (complex) surface state conductivity $\sigma_{2D}$ of the hyperbolic metal via $\sigma_{2D} = \frac{i\omega(\varepsilon_0 + \varepsilon_2)}{2\pi q_{2D}} = \frac{i\omega\kappa}{2\pi q_{2D}}$. Inside the hyperbolic regime of ZrSiSe, $\varepsilon_1$ is much larger than the dielectric function of the environment ($\varepsilon_0 = 1$, Air and $\varepsilon_2 = 1.94$, SiO$_2$). At $\omega = 6250\ cm^{-1}$ and considering only the real part, $\varepsilon_1 = \sqrt{\varepsilon_{ab}}\sqrt{\varepsilon_c} \approx 8.6i$ and therefore $i\frac{\varepsilon_j}{\varepsilon_1} \ll 1$ for $j = 0, 1$. Assuming $q_{2D} \ll q$, we can then approximate $\arctan\left[i\frac{\varepsilon_j}{\varepsilon_1}\left(1 - \frac{2q}{q_{2D}}\right)\right] \approx i\frac{\varepsilon_j}{\varepsilon_1}\left(1 - \frac{2q}{q_{2D}}\right)$ for $j = 0, 1$. Equation (S3) can then be simplified as:

$$q_l(\omega) = \frac{i\sqrt{\varepsilon_c}}{d\sqrt{\varepsilon_{ab}}} \left[ \pi l + i \frac{\varepsilon_0 + \varepsilon_2}{\sqrt{\varepsilon_{ab}}\sqrt{\varepsilon_c}} \left(1 - \frac{2q}{q_{2D}}\right) \right] \quad (S4)$$

and in the lossless limit (Re $\sigma_{2D} = 0$) we obtain:

$$q_l(\omega) = \frac{\pi l \sqrt{\varepsilon_c}\sqrt{|\varepsilon_{ab}|} + 2\kappa}{d|\varepsilon_{ab}| + 8\pi \operatorname{Im}(\sigma_{2D})/\omega} \quad (S5)$$

The influence of the surface state metallicity will be parameterized by its complex sheet conductivity $\sigma_{2D}$. If the imaginary part of $\sigma_{2D}$ at a given frequency is negative (positive), it will modify the HPP momentum to a larger (smaller) value, corresponding to enhanced (reduced) screening on the HPP dispersion.

Section S5: Angular and thickness dependence of hyperbolic plasmons near sample edges

In Fig. S12, we show the complete frequency-dependent near-field phase data ($\phi_4$) for the 20 nm thin ZrSiSe crystal on the SiO$_2$/Si substrate. Dashed lines indicate the paths along which we extract the line profiles. The phase derivative line profile ($\frac{d\phi_4}{dr}$) and the corresponding simulation are shown in Fig. S13. Circles and triangles marks the real-space features that correspond to the principal ($q_0$) and higher-order ($q_1$) HPP modes. In Fig. S14 we also show the full near-field amplitude ($S_4$) and phase ($\phi_4$) line profiles along the dashed lines in Fig. S12. The Fourier transform of the complex signal $S_4 e^{i\phi_4}$ are also shown in the right panel of Fig. S14 from $\omega = 5000$ cm$^{-1}$ to 8333 cm$^{-1}$. Blue and orange symbols correspond to the principal ($q_0$) and higher-order ($q_1$) HPP modes and are consistent with the momenta extracted through line profile modeling in Fig. S13, as shown in the main text. The Fourier transform between 5000 cm$^{-1}$ and 5780 cm$^{-1}$ are also indicative another higher-order mode ($q_2$, green symbols) predicted by the Im $r_p$ calculations.

The observed HPPs near sample edges are apparently angular dependent, as seen in Fig. S12 where the fringes in $\phi_4$ are more pronounced on the right edge than at the left edge. Such angular dependence is contained within our quasistatic model (see Sec. S7) and stems from the polarization of the sample by the external field. This effect is indeed reflected in the modeled near-field image shown in Fig. S19. Importantly, the entire phase simulation image shown in Fig. S19 is generated with the polariton wavelengths obtained from the derivative line profile modelling in Fig. S13. The good agreement between the experiment and simulation on both edges of the crystal further confirms the accuracy of the extracted polariton wavelengths through line profile modelling.

As with the antenna launched HPPs, the tip-launched modes also show distinct thickness dependence that can be directly compared with the maximum of Im ($r_p$) calculated based on experimental dielectric functions (Fig. S1 and Fig. S5). In Fig. S15 we show the topography and corresponding near-field phase images ($\omega = 8333 \, cm^{-1}$) of a multi-terraced ZrSiSe crystal on SiO$_2$/Si substrate. The thin flakes ranges from 24 to 122 nm in thickness and the phase linecut (Fig. S15C) shows an increase in fringe periodicities with increasing sample thickness. In particular, the distance between the first peak and the first dip ($t$) is approximately 0.13 times of modeled plasmon wavelength ($t \approx 0.13\lambda_p$). Such direct extraction of polariton wavelength has been utlized before in monolayer hBN (*12*) and serve as a quick estimate of the HPP wavelength in ZrSiSe. In the inset of Fig. S15C, we plot the extracted HPP momentum ($q_0 = \frac{2\pi}{\lambda_p}$) estimated based on the distance of the first peak and the first dip for various thicknesses of ZrSiSe. The data points are normalized to the momentum of the free-space light and agrees well with the calculated maxima of Im ($r_p$) (red curve).

Section S6: Modeling the near-field signal near antenna edges

To model the spatial profile of the signal near the edge of the gold disk, we develop an approximate solution for the scattered field created by a conducting disk, including the effects of diffraction. The basis of this approximation is Sommerfeld's solution to the famous problem of diffraction by a perfectly conducting screen (*12*). Below, we review this solution and use it to construct an approximate solution for a metallic disk covered by a thin optically hyperbolic film. Consider first a wave, incident at an angle $\alpha$ with respect to the plane with no component parallel to the edge of the conducting screen ($\beta = 0$), which we denote as the $y$-direction (Fig. S16 left). For concreteness, we first consider the case of a magnetic field $\boldsymbol{H} = H_y\hat{\boldsymbol{y}}$, where the scattered magnetic field has only one component along the $y$-direction, $H_y = U^\perp(x,z)$. The scattered magnetic field can be expressed through Fresnel diffraction integrals $F(z)$:

$$U^\perp(x,z;k) = U_0(x)\left(\frac{e^{ikz\sin\alpha + \frac{i\pi}{4}}}{\sqrt{\pi}}\left(F(\eta_+) + \frac{e^{-ikz\sin\alpha + \frac{i\pi}{4}}}{\sqrt{\pi}}F(\eta_-)\right) - i\sin(kz\sin\alpha)\right) \quad (S6)$$

where $F(z) = \int_0^z e^{-i\kappa^2}d\kappa$, $\eta_\pm = \sqrt{2kr}\cos\frac{\phi\mp\alpha}{2}$, and $U_0(x) = E_0 e^{ikx\cos\alpha}$. Here $k$ is the free-space photon wavevector, and $(r, \phi)$ represents polar coordinates in the $xz$-plane with $\tan\phi = \frac{z}{x}$ and $r = \sqrt{x^2 + z^2}$. For the other, orthogonal polarization with the incident electric field $\boldsymbol{E} = E_y\hat{\boldsymbol{y}}$, one obtains the second solution for $E_y = U^\parallel(x,z)$:

$$U^\parallel(x,z;k) = U_0(x)\left(\frac{e^{ikz\sin\alpha + \frac{i\pi}{4}}}{\sqrt{\pi}}\left(F(\eta_+) - \frac{e^{-ikz\sin\alpha + \frac{i\pi}{4}}}{\sqrt{\pi}}F(\eta_-)\right) - \cos(kz\sin\alpha)\right) \quad (S7)$$

An arbitrary incidence angle relative to the edge can be accomplished by introducing an angle $\beta$, understood as a latitude relative to the *y*-axis, shown in Fig. S16. The angles $\alpha, \beta$ are related to the incidence angles $\theta, \phi$ of a spherical polar coordinate system by the relations:
$$\cos\alpha \cos\beta = \cos\psi \sin\theta$$
$$\sin\beta = \sin\psi \sin\theta$$
$$\sin\alpha \cos\beta = \cos\theta$$
The z-component of the scattered electric field for an incident *p*-polarized light can then be decomposed into the polarizations of the fundamental solutions $U^\perp, U^\parallel$ (13), yielding:
$$E_z^{sca}(x,y,z;k) = e^{iky\sin\beta}\left[A\frac{i}{k}\frac{\partial}{\partial x}\left(U^\perp(x,z;k\cos\beta)\right) + B\frac{i\sin\beta}{k}\frac{\partial}{\partial z}\left(U^\parallel(x,z;k\cos\beta)\right)\right] \quad (S8)$$
The coefficients $A, B$ arise from the decomposition of the polarization of the incident wave into components parallel and perpendicular to the edge of the screen and depend only on the polar and in-plane incidence angles $\alpha, \beta$. We can then construct an approximate solution for a disk by solving for several angles $\psi$ and plotting the diffraction pattern produced for each angle, with the out-of-plane component $E_z$ plotted in Fig. S17a at a frequency of $\omega = 6600\ cm^{-1}$.

To check the validity of this approximation, we used the COMSOL package to simulate the scattered field distribution produced by a plane wave whose magnetic field was polarized parallel to the disk. This numerical approach was necessitated by the large free-space wavelength, which is comparable to the size of the metallic disk, invalidating the quasistatic approximation typically used in the modeling of the SNOM signal. The disk was included by implementing a perfectly conducting boundary condition on the surface of the disk inside of a physical domain of dimension $4\ \mu m \times 4\ \mu m \times 2\ \mu m$ padded with perfectly matched layers of thickness 500 nm at each edge of the domain. A scattering boundary condition was implemented at the edge of the physical domain, and only the scattered field was extracted. The result of this simulation is plotted in Fig. S17b. The agreement between the approximation and the numerical solution is expected to hold only near the edge of the disk, which contains the crucial feature, namely a divergence of the field due to a sharp edge. The angular intensity distribution around the circumference of the disk is also captured by the approximate model, which can then be modified to account for the effect of the hyperbolic medium.

The introduction of the sample will bring with it the hyperbolic modes and modify the scattered field. The multiple branches of the polariton dispersion observed are derived by computing the poles in the reflection coefficient $r_p(q, \omega)$ in the absence of losses. In a realistic system with finite loss, the dispersion is instead dictated by the maxima in Im $r_p(q, \omega)$. We consider a three-layer system consisting of vacuum, sample and substrate, labeled as medium 0, 1 and 2, respectively. The divergence of $r_p(q, \omega)$ happens at a discrete set of values satisfying the condition:
$$2\pi l + \psi_{01} + \psi_{21} = 2k_1^z d \quad (S9)$$

for a medium of thickness $d$. The phase shifts $\psi_{01}, \psi_{21}$ can be expressed in terms of reflection coefficients at the top and bottom interfaces, $r_{01} = e^{i\psi_{01}}$ and $r_{21} = e^{i\psi_{21}}$, respectively. The reflection coefficients $r_{ij}$ at the interfaces are given by:

$$r_{ij}(q) = \frac{\frac{\varepsilon_j^\perp}{k_j^z} - \frac{\varepsilon_i^\perp}{k_i^z}}{\frac{\varepsilon_j^\perp}{k_j^z} + \frac{\varepsilon_i^\perp}{k_i^z}} \tag{S10}$$

where the $z$-component of the wavevector $k_i^z$ of a $p$-polarized light in each medium is given by:

$$k_i^z(q) = \sqrt{\varepsilon_i^\perp}\sqrt{\frac{\omega^2}{c^2} - \frac{q^2}{\varepsilon_i^z}}, \qquad \operatorname{Im} k_i^z > 0 \tag{S11}$$

In the hyperbolic regime $(q)\omega/c$, $k_1^z$ is predominantly real, so the solutions of Eqn. (S9) are not confined to a surface but can exist within the bulk of the sample. A closed-form solution for the dispersion can only be obtained within the quasistatic approximation ($c \to \infty$). In that case, the reflection coefficients of Eqn. (S10) become independent of $q$ and reduce to:

$$\beta_{ij} = \frac{\sqrt{\varepsilon_j^\perp}\sqrt{\varepsilon_j^z} - \sqrt{\varepsilon_i^\perp}\sqrt{\varepsilon_i^z}}{\sqrt{\varepsilon_j^\perp}\sqrt{\varepsilon_j^z} + \sqrt{\varepsilon_i^\perp}\sqrt{\varepsilon_i^z}} \tag{S12}$$

The original transcendental equation Eqn. (9) reduces to a linear equation for the dispersion of each mode $q_l$. In the particular case of launching by a conducting metallic edge, the observed fringes in real space can be understood as a beating between the various modes $q_l$ in momentum space (*14*), giving a fringe spacing of:

$$\lambda_p = \frac{2\pi}{\Delta q_l} \approx -2id\frac{\sqrt{\varepsilon_1^\perp}}{\sqrt{\varepsilon_1^z}} \tag{S13}$$

with the last equality holding in the quasistatic limit.

Having previously obtained a solution for the field created in vacuum by a screen, this expression can be used as a building block to construct an approximate solution to the field produced by the system of the disk, sample, and substrate. Since the polariton wavelength ($\lambda_p$) is an order of magnitude smaller than the free-space photon wavelength $\lambda_0$, near the edge we expect a quasistatic approximation to be valid, permitting the use of an image method to introduce a sample (*14*). Using the field from Eqn. (S8), we introduce an equidistant series of images, as in the solution for the static field of a dielectric film between two media. We take the infinitely thin disk as the source of this static field, situated at the interface of media 1 and 2, that is, below the ZrSiSe layer. For an optically anisotropic material, the thickness of the film is further modified by the ratio of in-plane and $z$-axis dielectric function: $\frac{\sqrt{\varepsilon^t}}{\sqrt{\varepsilon^z}}$. The scattered near-field signal ($S$) can be approximated as the $z$-components of the field $E_z^S$ obtained from the diffraction problem and the reflection coefficients $\beta_{ij}$ from Eqn. (S12):

$$S(x,y,z) = (1 - \beta_{01})E_z^{sca}(x,y,z+h) + \beta_{21}(1+\beta_{01})\sum_{n=1}^{\infty}\beta_{01}^n\beta_{21}^n E_z^{sca}\left(x,y,(2n+1)d\frac{\sqrt{\varepsilon^t}}{\sqrt{\varepsilon^z}}+h\right) \quad (S14)$$

To include the effects of demodulation, we compute the field at a discrete set of points above the sample:

$$h(t) = h_0 + \Delta h(1 - \cos n\Omega t) \quad (S15)$$

to obtain the complex signal $\tilde{s}_n = S_n e^{i\phi_n}$. Here $\Omega$ is the tip-tapping frequency and we used tapping amplitude $\Delta h = 50\ nm$ and minimum position $h_0 = 5\ nm$. The demodulated scattering amplitudes ($n = 3$) computed at a few different frequencies are shown in Fig. S18.

Section S7: Simulation of plasmonic fringes near the sample edge.

The input parameter into the model is a complex wavevector $Q = q_p(1 + i\gamma)$, where $q_p = \frac{2\pi}{\lambda_p}$ and $\gamma$ represents the dimensionless damping coefficient of the polariton mode. For very thin layers ($d \ll \lambda_p$), one can approximate the sample by a two-dimensional conducting layer with an effective sheet conductivity

$$\sigma_{eff} = \frac{i\omega\kappa}{2\pi Q}, \quad (S16)$$

with $\kappa = \frac{\varepsilon_0 + \varepsilon_2}{2}$ being the average permittivity of the surrounding media. The sample is modeled by a two-dimensional strip of width L at the boundary of two half-spaces with permittivities $\varepsilon_0 = 1$ (air) and $\varepsilon_2 = 1.94$ (SiO$_2$). The SNOM signal is taken to be proportional to the induced dipole moment on the probe. The scanning probe is modeled by a spheroid, with radius of curvature a = 40 nm and length L = 1400 nm. We compute the charge distribution $n_i$ in the sample induced by the probe. This quantity can be found by combining Gauss's law with the charge continuity equation, giving

$$\Phi(\mathbf{r}) - V(\mathbf{r}) * n_i(\mathbf{r}) = \Phi_{ext}(\mathbf{r}), \quad (S17)$$

with the operation $A * B$ denoting convolution, $\Phi(\mathbf{r})$ being the full potential, and $V(\mathbf{r})$ being the in-plane Coulomb potential. The external field is taken to be constant with an incidence angle $\theta = 60°$ from the surface normal. Using the translation symmetry of the problem in the lateral direction, Eqn. (S17) is reduced to a one-dimensional integral equation. Replacing the derivatives by finite differences further simplifies Eqn. (S17) to a matrix inversion. The SNOM signal is then found by computing the dipole moment induced on the probe by the distribution $n_i(\mathbf{r})$. The complex signal (S) is calculated for a range of tip positions from each quantity $n_i(\mathbf{r})$ and then demodulated to the 4th harmonic in order to compare with the experimental line profiles shown in Fig. 3D in the main text and Fig. S13.

To simulate the near-field phase-contrast of triangular-shaped samples (Fig. S19), we created the two-dimensional image by solving Eqn. (S17) for a semi-infinite conducting sheet near a sample edge. We obtained two different solutions for the two cases of the in-plane component of the external field being parallel and anti-parallel to the sample edge. These two solutions were used for each edge of the triangular flake separately, and the intermediate region was interpolated between the edges. This approach is valid provided that the width of the sample at that point $L \gg \operatorname{Im} Q$, see Eqn. (S16).

Section S8: Survey of electronic loss in plasmonic and excitonic hyperbolic materials
In this section, we list the reflectance, dielectric function and optical conductivities ($\sigma(\omega) = \sigma_1 + i\sigma_2$) of various plasmonic and excitonic hyperbolic materials reported in the literature. To quantify the electronic loss, the ratio $\frac{\sigma_2}{\sigma_1}$ is calculated based on the reported or extracted optical conductivities.

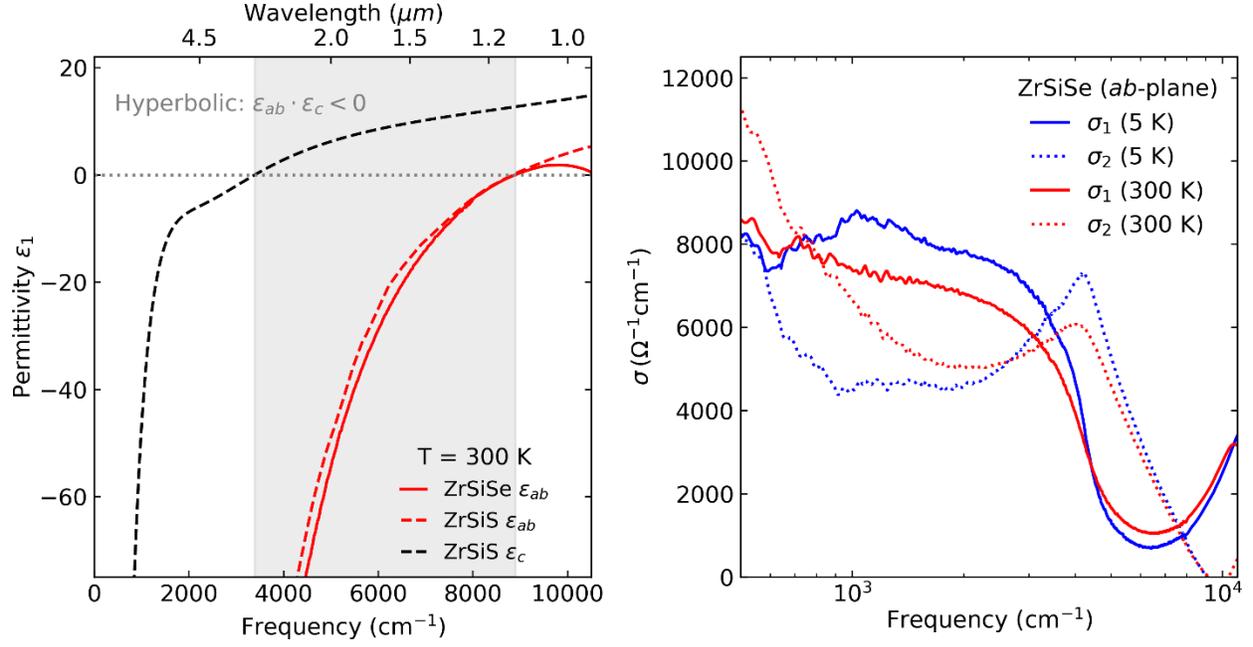

**Fig. S1. Optical conductivities of ZrSiS and ZrSiSe.** (Left) Real part of the dielectric function of ZrSiS (dashed lines) and ZrSiSe (solid line). The gray-shaded region indicates the frequency range where ZrSiS is hyperbolic. (Right) *ab*-plane optical conductivities of ZrSiSe at 300 K and 5 K. Solid and dotted lines represent the real and imaginary parts of $\boldsymbol{\sigma(\omega) = \sigma_1 + i\sigma_2}$, respectively.

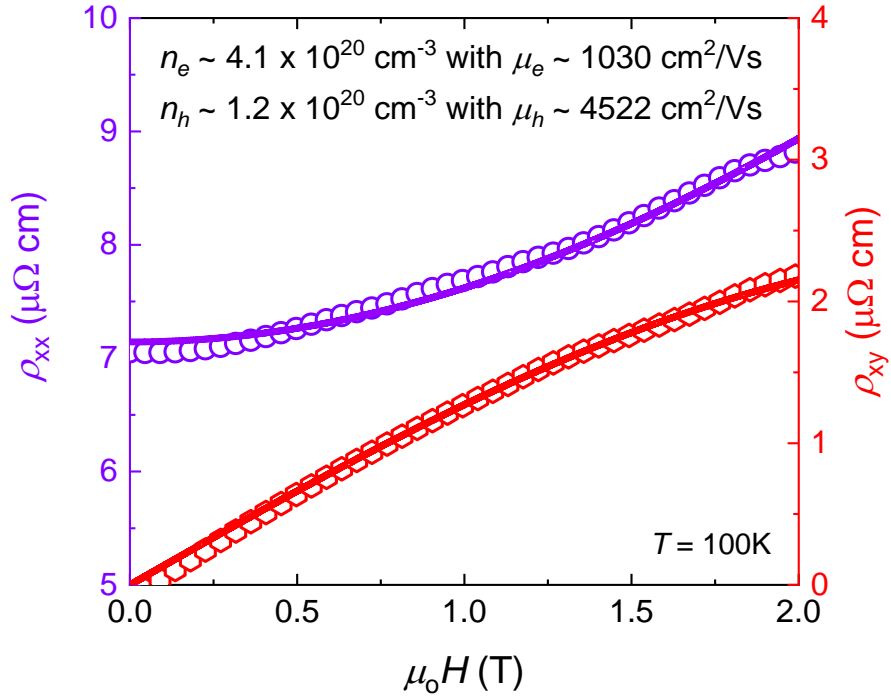

**Fig. S2**. **Magnetotransport measurements and fitting.** Two-band model fits of longitudinal ($\rho_{xx}$) and transverse ($\rho_{xy}$) resistivity for ZrSiSe at 100 K.

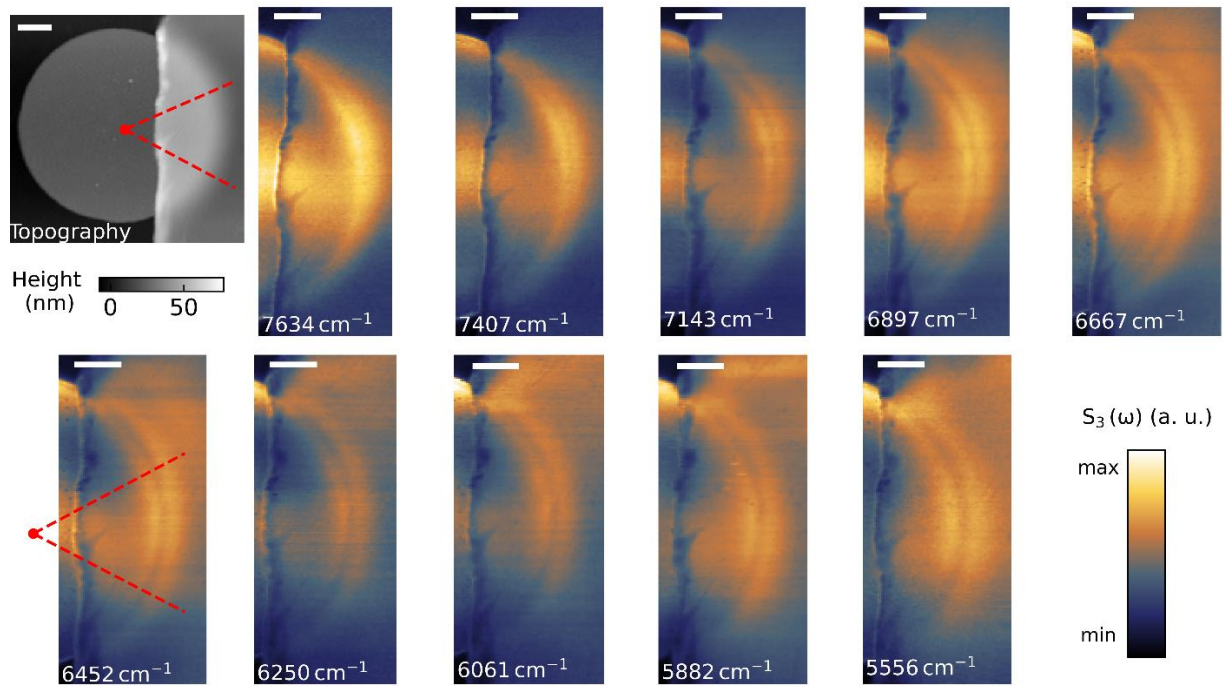

**Fig. S3**. **Frequency dependent near-field data.** Topography and frequency-dependent near-field scattering amplitude data ($S_3$) of the 26 nm ZrSiSe sample on a gold disk antenna. Scale bars in all panels are 300 nm. Red dashed lines indicate the sector region used to average the line profiles of $S_3$ along the perimeter of the disk antenna and are kept the same for all frequencies.

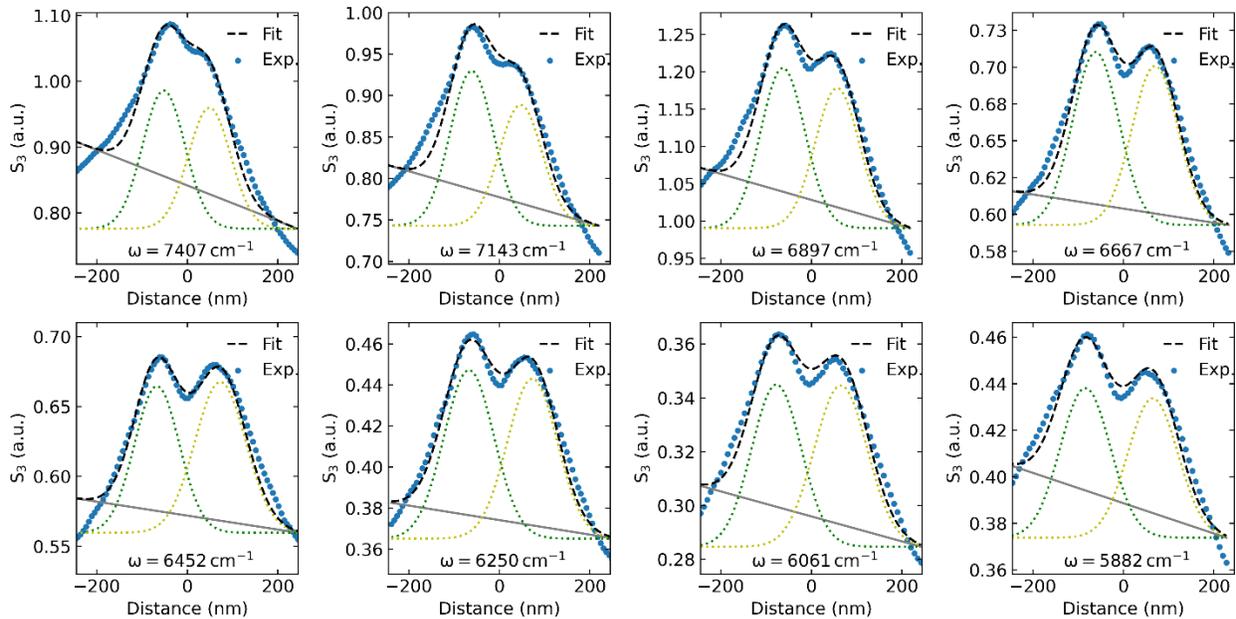

**Fig. S4**. **Extracted line profiles and fitting.** Frequency-dependent line profiles in the sample region from Fig.S3 are shown as blue dots. The extracted line profiles are fitted with two Gaussian profiles (green and yellow dotted lines) and linear backgrounds (gray solid lines). Black dashed lines are the sum of the Gaussians and the background, showing good agreement with the experiment.

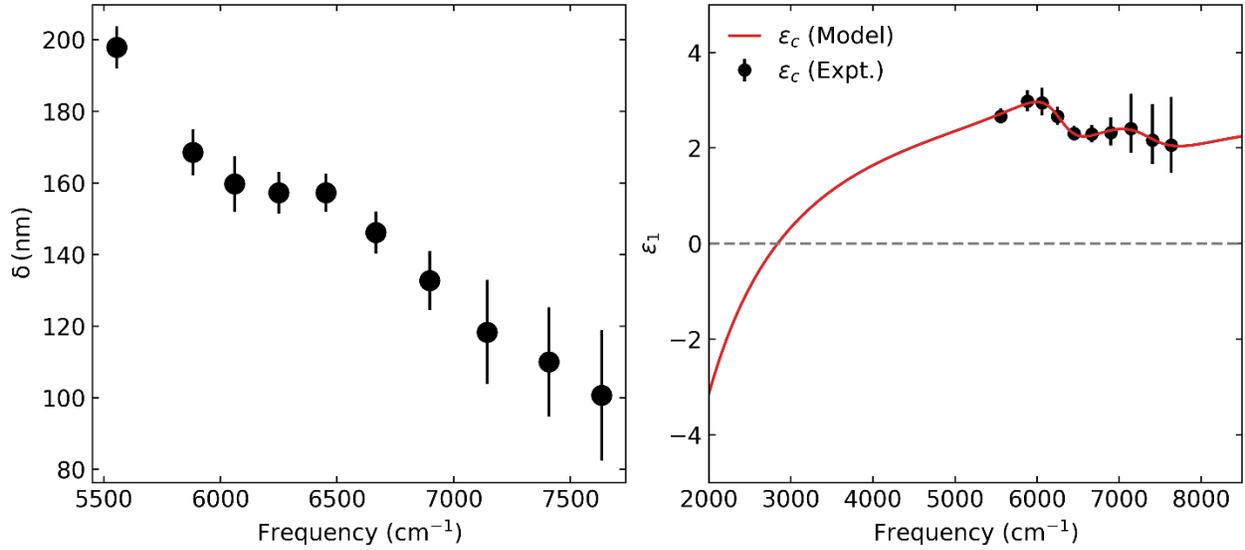

**Fig. S5. Double-ring separation and Drude-Lorentz model of the *c*-axis dielectric function for ZrSiSe.** (Left) Experimental peak separation $\delta(\omega)$ obtained from the fitting in Fig. S4 and the slope correction discussed in the Materials and Methods section. (Right) Drude-Lorentz model fitting of the *c*-axis dielectric function data (black dots), obtained through the antenna launching experiment.

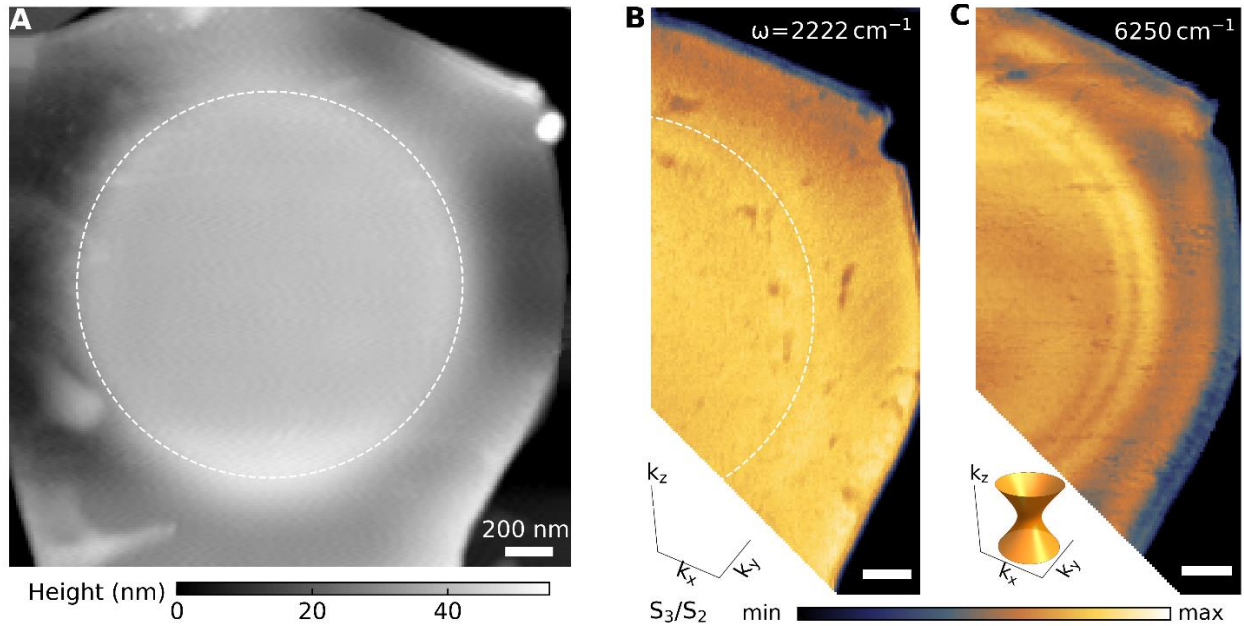

**Fig. S6. Hyperbolic ray launching experiment for fully covered Au antenna.**
(**A**), Topography image of a 28 nm ZrSiSe crystal on top of a 2 μm wide Au circular antenna. The normalized near-field amplitudes ($\frac{S_3}{S_2}$) outside ($\omega = 2222$ cm$^{-1}$) and inside ($\omega = 6250$ cm$^{-1}$) the hyperbolic regime are shown in panel (**B**) and (**C**), respectively. Insets in panels **B** and **C** are the schematic of the isofrequency surfaces. Dashed lines in **A** and **B** indicate the boundary of the underlying Au antenna. Scale bars in all panels are 200 nm.

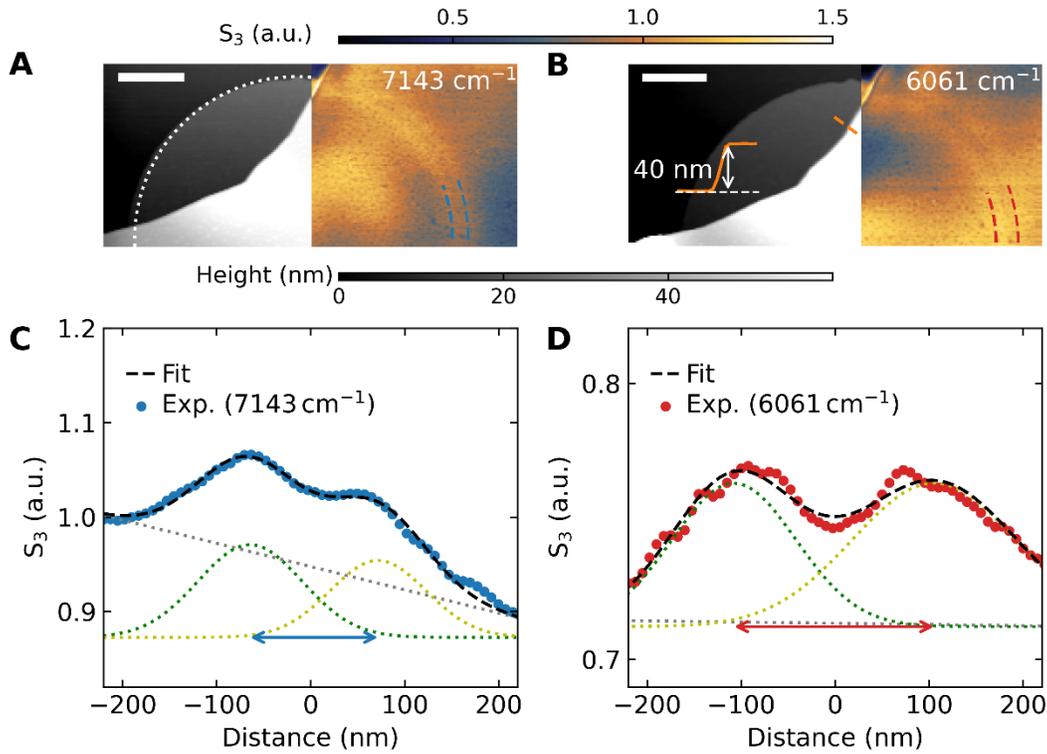

**Fig. S7. Hyperbolic ray launching data and fitting of a 40 nm ZrSiSe crystal on circular gold antenna**. Topography (left) and the near-field amplitude data (right) collected at (**A**) ω = 7143 cm$^{-1}$ and (**B**) ω = 6061 cm$^{-1}$. The white dotted line indicates the boundary of gold. Blue and red dashed lines indicate the double-ring features at 7143 cm$^{-1}$ and 6061 cm$^{-1}$, respectively. Inset in **B** is a topography linecut along the orange dashed line. Scalebars are 500 nm. The averaged line profiles of the near-field amplitude along the perimeter of the circular antenna are shown for (**C**) ω = 7143 cm$^{-1}$ and (**D**) ω = 6061 cm$^{-1}$. The line profiles are fitted with two Gaussian profiles (green and yellow dotted lines) and a linear background (gray dotted line).

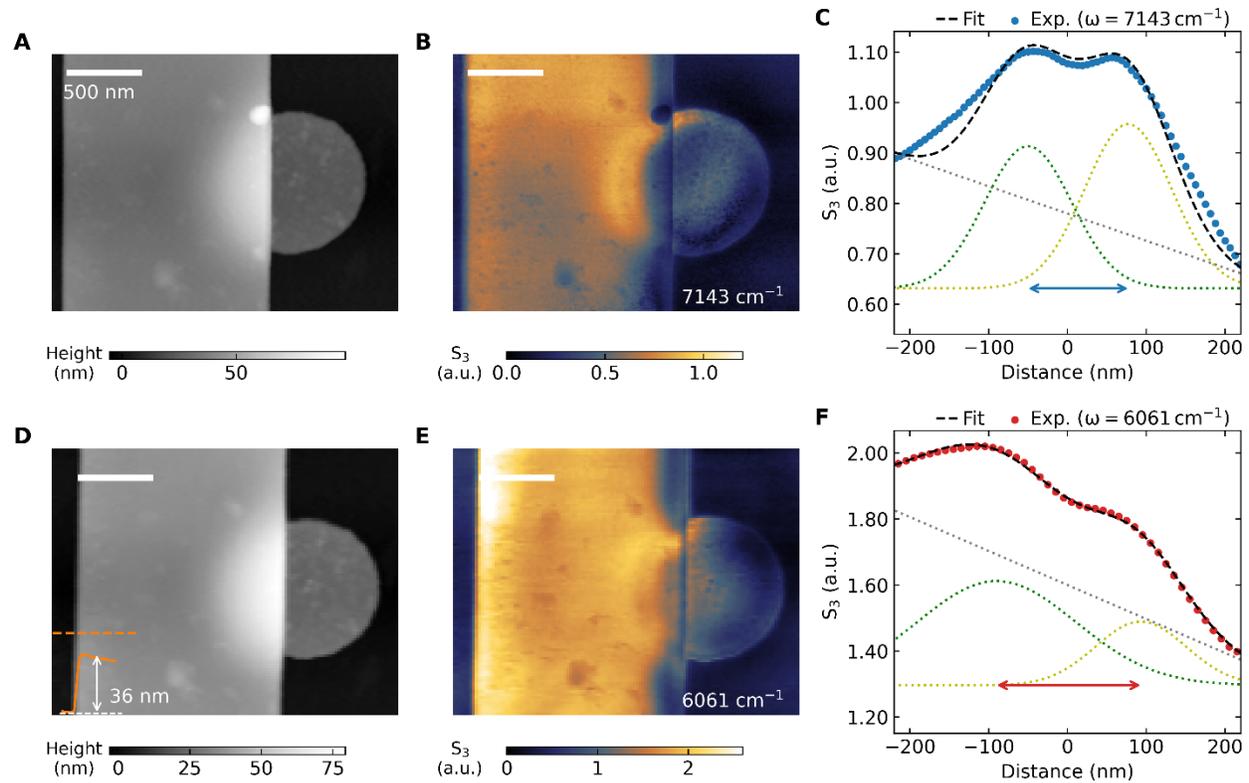

**Fig. S8. Hyperbolic ray launching data and fitting of a 36 nm ZrSiSe crystal on circular gold antenna**. Topography and the corresponding near-field amplitude data ($S_3$) at (**A**, **B**) $\omega = 7143\ cm^{-1}$ and (**D**, **E**) $\omega = 6061\ cm^{-1}$. Inset in panel **D** is the topography line profile along the orange dashed line. The extracted near-field amplitude line profiles on the sample along the perimeter of the antenna are shown for $7143\ cm^{-1}$ and $6061\ cm^{-1}$ in **C** and **F**, respectively. The line profiles are fitted with Gaussian functions (green and yellow dotted lines) and a linear background (grey dotted line). Scale bars in panels **A**-**D** are 500 nm.

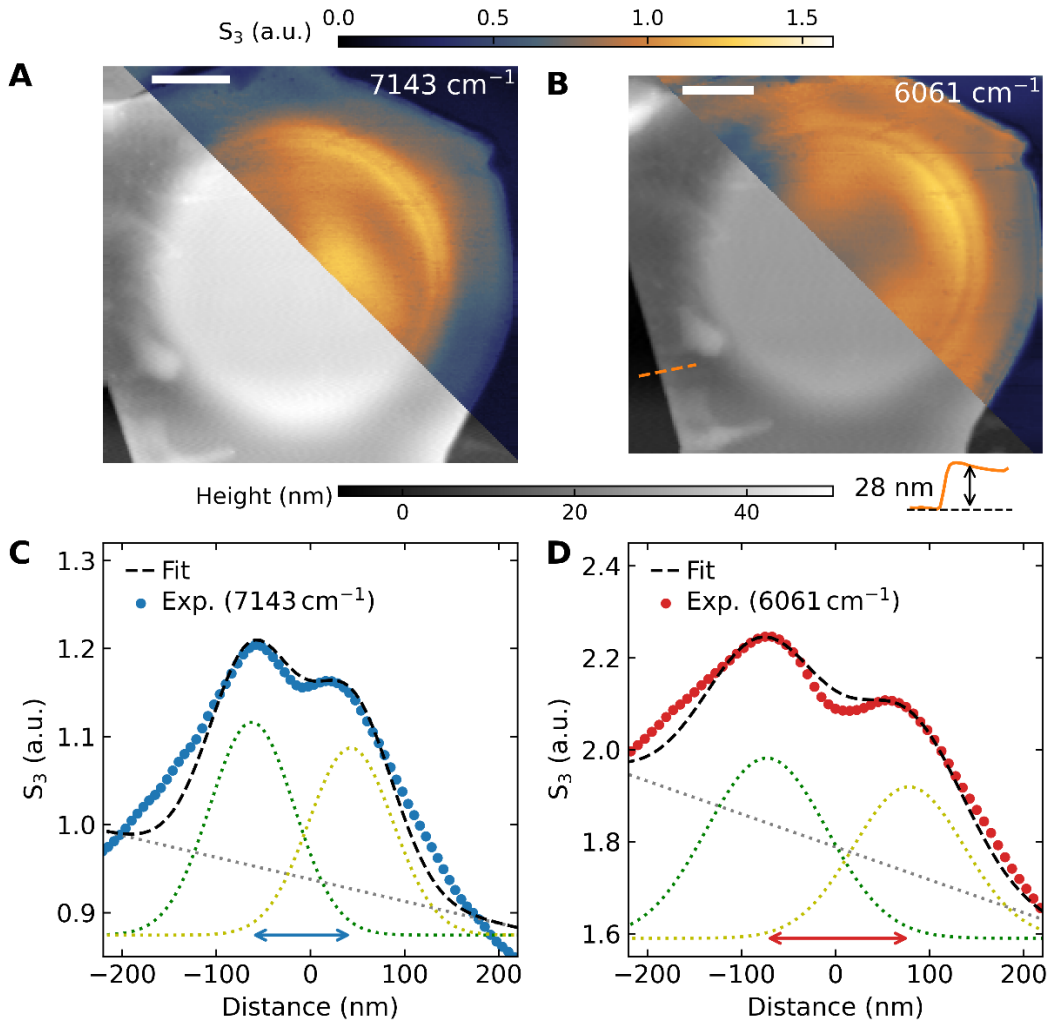

**Fig. S9. Hyperbolic ray launching data and fitting of a 28 nm ZrSiSe crystal on circular gold antenna**. Topography (bottom) and the near-field amplitude data (top) at (**A**) $\omega = 7143\ cm^{-1}$ and (**B**) $\omega = 6061\ cm^{-1}$. The extracted near-field amplitude line profiles on the sample along the perimeter of the antenna are shown for $7143\ cm^{-1}$ and $6061\ cm^{-1}$ in **C** and **D**, respectively. The line profiles are fitted with Gaussian functions (green and yellow dotted lines) and a linear background (grey dotted line). Scale bars in panels **A**,**B** are 500 nm.

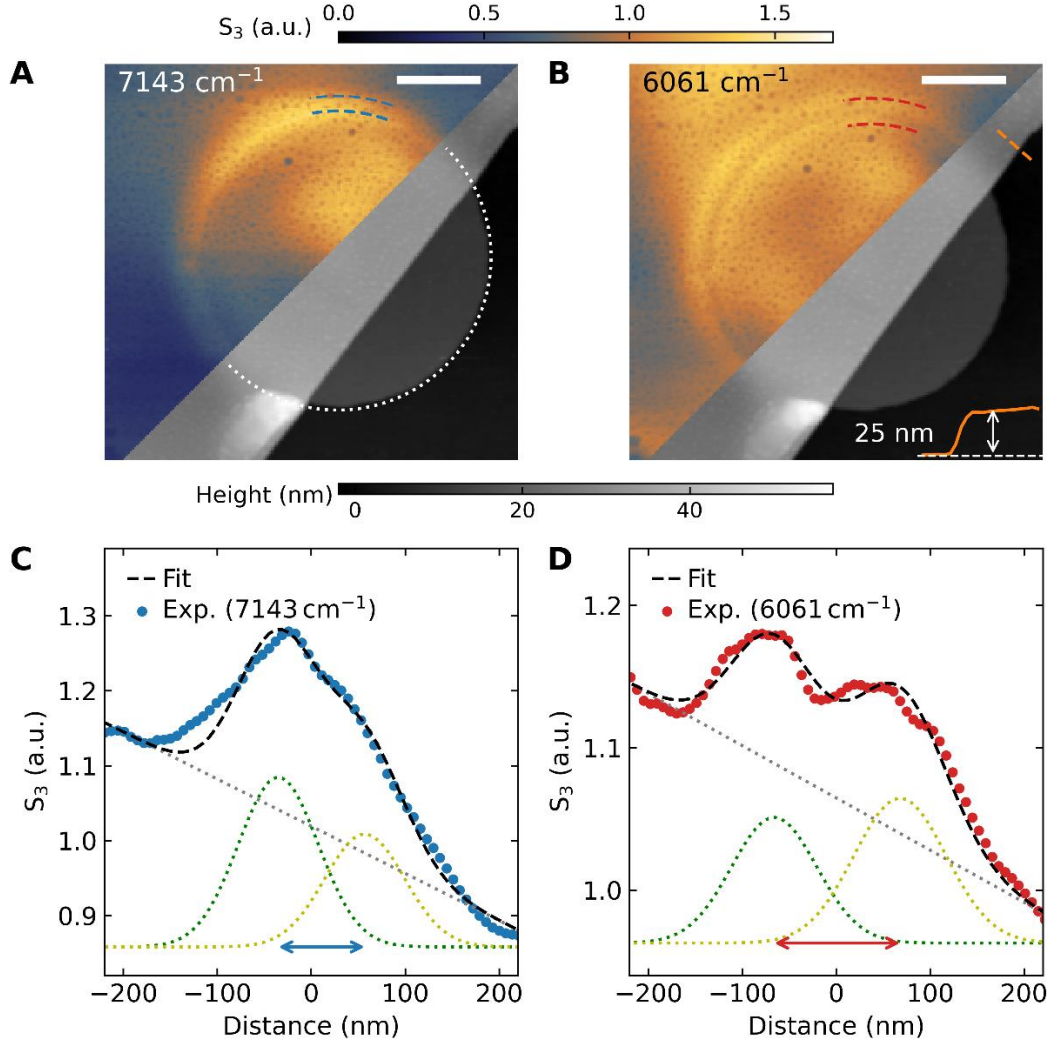

**Fig. S10. Hyperbolic ray launching data and fitting of a 25 nm ZrSiSe crystal on circular gold antenna**. Topography (bottom) and the near-field amplitude data (top) collected at (**A**) $\omega = 7143$ cm$^{-1}$ and (**B**) $\omega = 6061$ cm$^{-1}$. The white dotted line indicates the boundary of gold. Blue and red dashed lines indicate the double-ring features at 7143 cm$^{-1}$ and 6061 cm$^{-1}$, respectively. Inset in **B** is a topography linecut along the orange dashed line. Scalebars are 500 nm. The averaged line profiles of the near-field amplitude along the perimeter of the circular antenna are shown for (**C**) $\omega = 7143$ cm$^{-1}$ and (**D**) $\omega = 6061$ cm$^{-1}$. The line profiles are fitted with two Gaussian profiles (green and yellow dotted lines) and a linear background (gray dotted line).

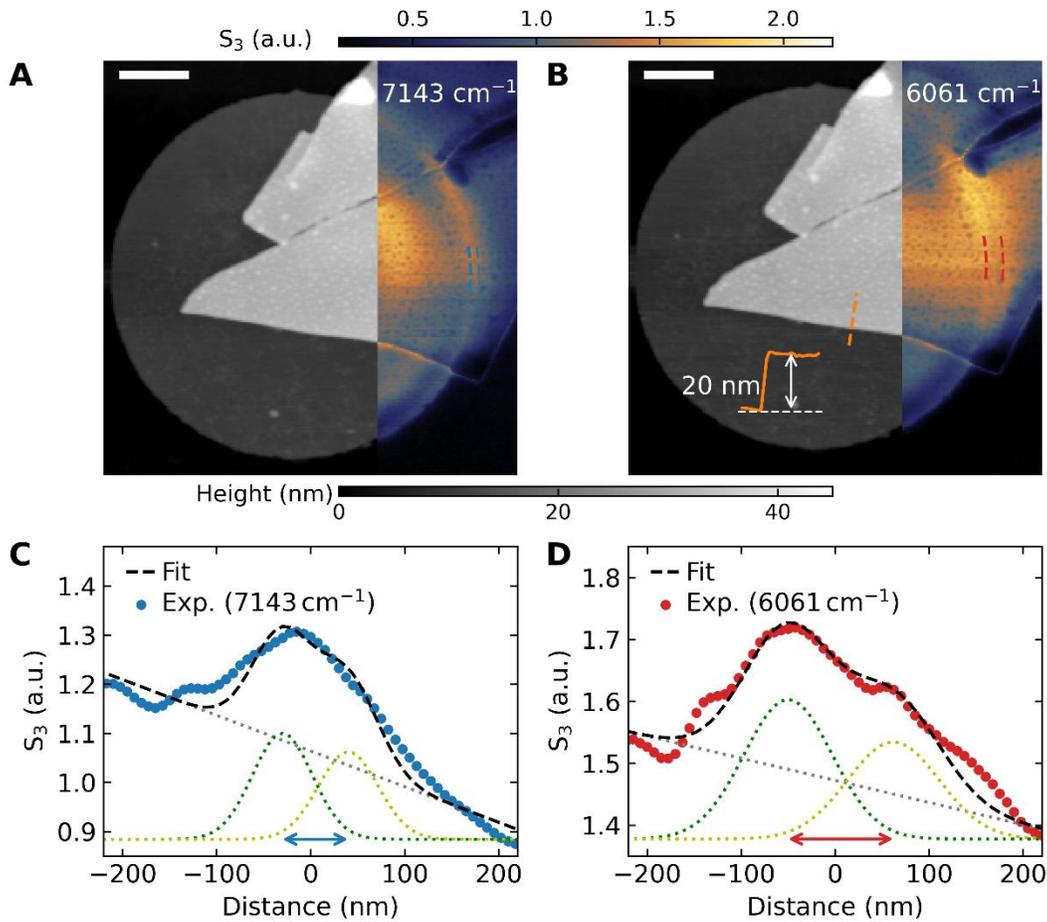

**Fig. S11. Hyperbolic ray launching data and fitting of a 20 nm ZrSiSe crystal on circular gold antenna**. Topography (left) and the corresponding near-field amplitude data (right) at (**A**) $\omega = 7143\ cm^{-1}$ and (**B**) $\omega = 6061\ cm^{-1}$. Inset in panel **B** is the topography line profile along the orange dashed line. The extracted near-field amplitude line profiles on the sample along the perimeter of the antenna are shown for $7143\ cm^{-1}$ and $6061\ cm^{-1}$ in **C** and **D**, respectively. The line profiles are fitted with Gaussian functions (green and yellow dotted lines) and a linear background (grey dotted line). Scale bars in **A**,**B** are 500 nm.

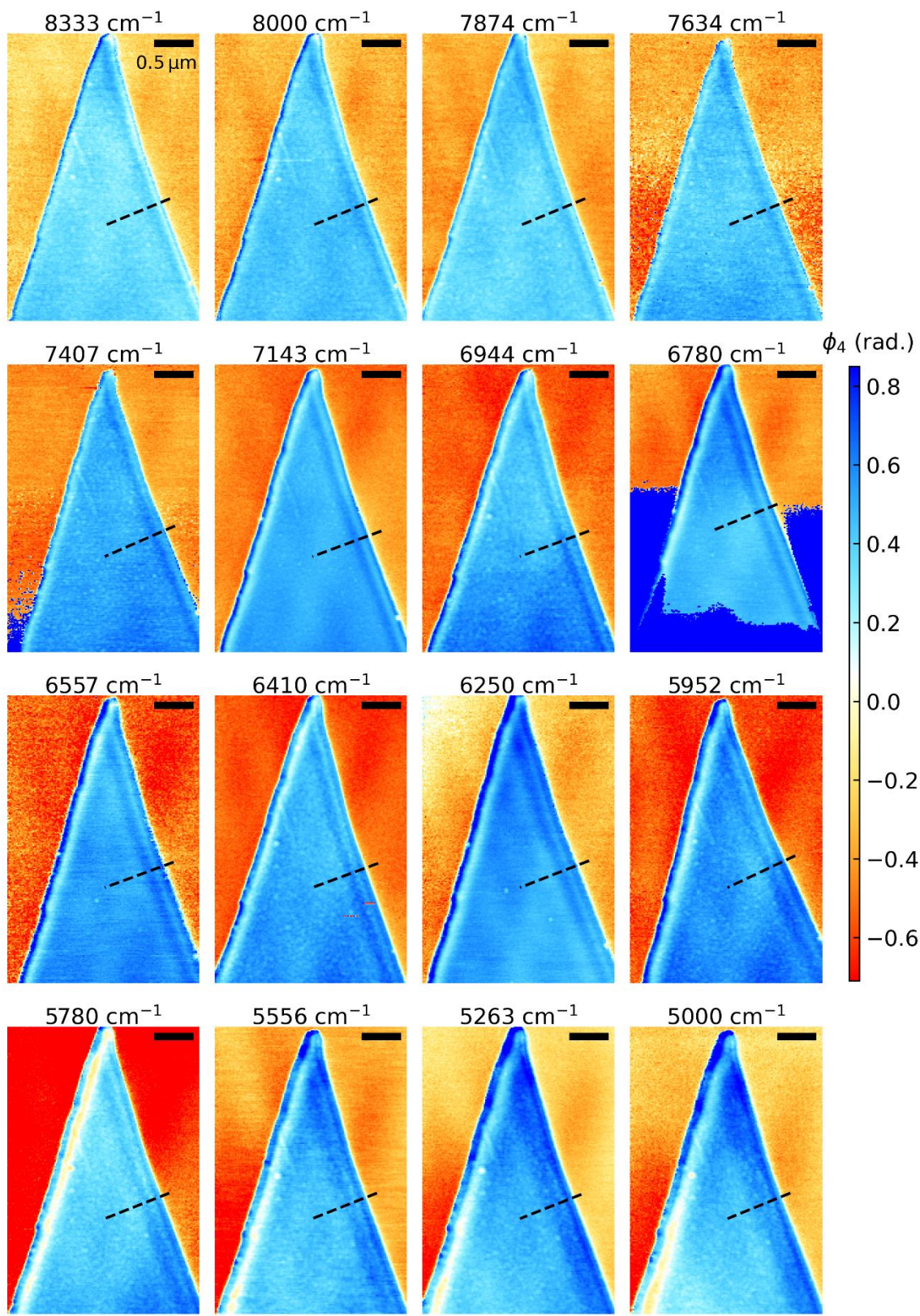

**Fig. S12. Edge launching near-field imaging data.** Frequency-dependent ($\omega = 8333 - 5000\ cm^{-1}$) near-field phase ($\phi_4$) for the 20 nm ZrSiSe on SiO$_2$/Si. Scale bars are 500 nm.

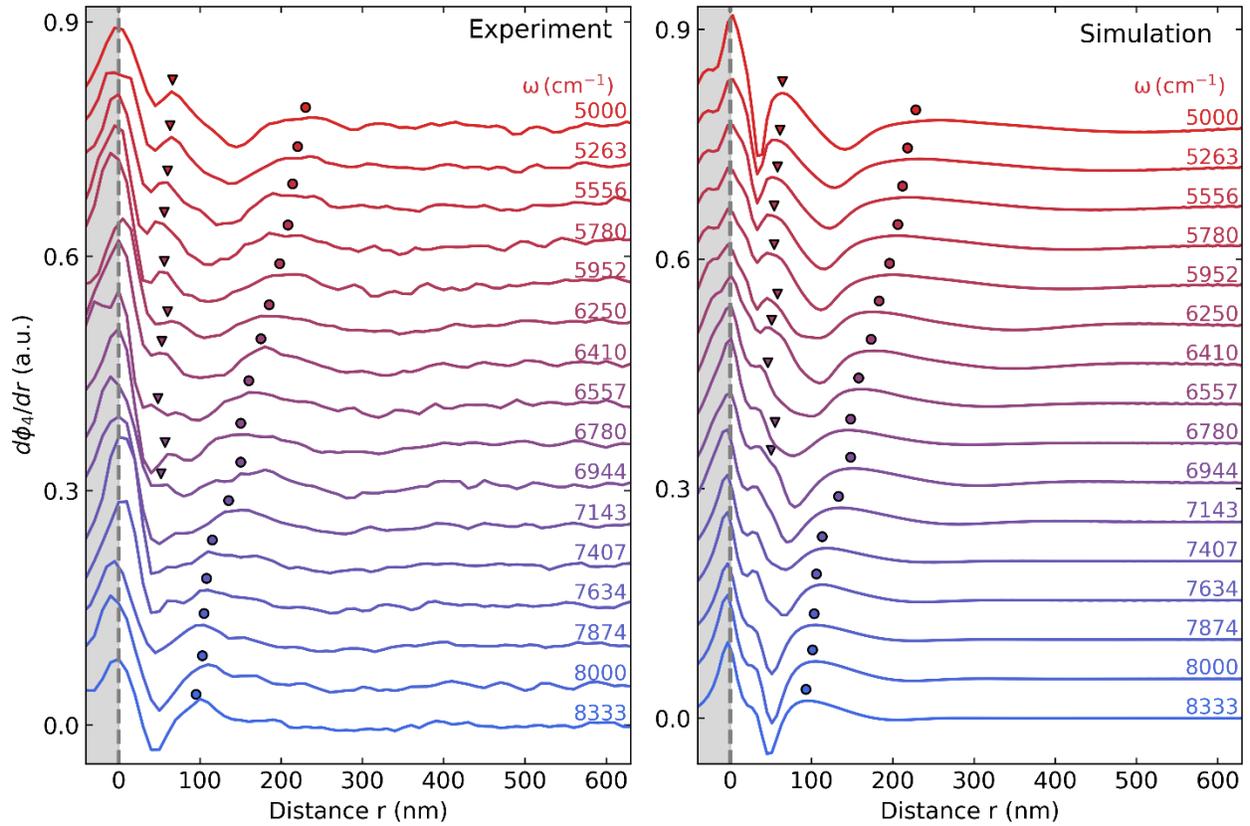

**Fig. S13. Experimental and simulated phase derivative line profiles near edges of ZrSiSe.**
(Left) Line profiles of near-field phase derivative ($\frac{d\phi_4}{dr}$) along the black dashed lines in Fig. S12 for the 20 nm ZrSiSe on SiO$_2$/Si substrate. (Right) Simulation of the phase derivative line profiles at corresponding frequencies. Colored circles and triangles mark the positions of the principal ($q_0$) and higher-order ($q_1$) hyperbolic plasmon polaritons, respectively.

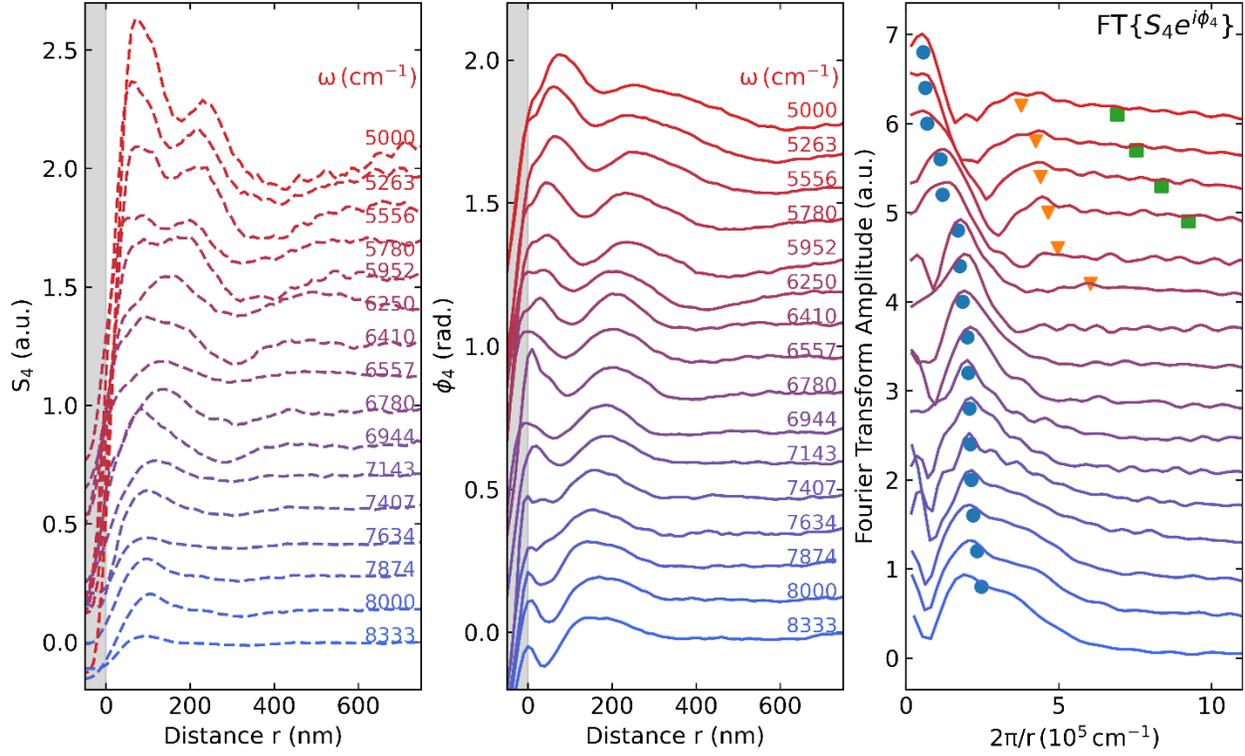

**Fig. S14. Amplitude and phase line profiles near edges of ZrSiSe and the corresponding Fourier transform amplitude.** (Left and Middle) Experimental line profiles of near-field amplitude ($S_4$) and phase ($\phi_4$) along the black dashed lines in Fig. S12 for the 20 nm ZrSiSe on SiO$_2$/Si. (Right) Fourier transform of the complex signal $S_4 e^{i\phi_4}$ from $\omega = 5000$ cm$^{-1}$ (red) to 8333 cm$^{-1}$ (blue). Blue, orange and green symbols represent the mometa of the principal ($q_0$) and higher-order hyperbolic plasmon polaritons modes ($q_1, q_2$), respectively.

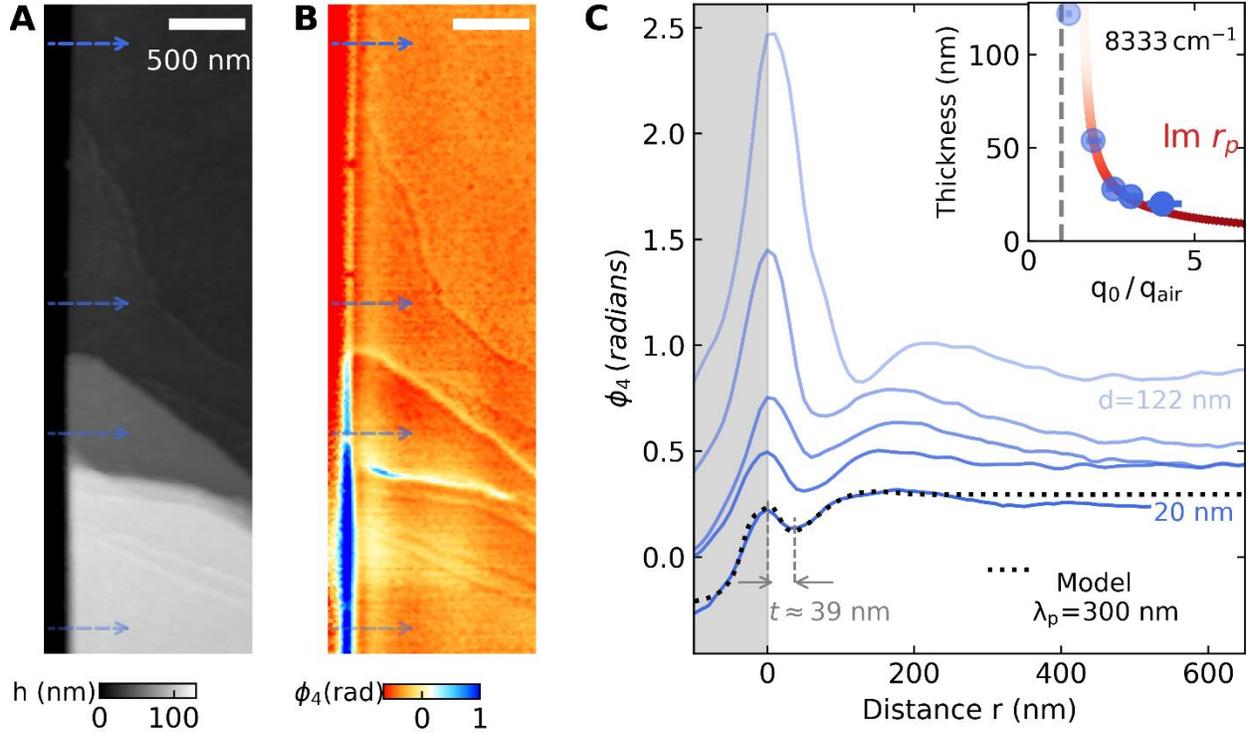

**Fig. S15. Thickness dependence of edge launched HPP thin ZrSiSe.** (**A**), Topography of a multi-terraced ZrSiSe crystal on SiO$_2$/Si substrate. (**B**), Near-field phase image ($\phi_4$) in the same region taken at $\omega = 8333\ cm^{-1}$. The four arrows from top to down corresponds to thickness of 24 nm, 28 nm, 50 nm and 122 nm, respectively. (**C**), Phase line profiles at various thicknesses along the arrow positions in panel (B), the line profile and model for 20 nm ZrSiSe are taken from Fig. S12 (right edge). Gray shaded region indicates the substrate. Black dotted line indicate the modeled line profile with plasmon wavelength ($\lambda_p$) of 300 nm. Inset depicts the thickness dependence of extracted (principal) HPP momentum $q_0 = \frac{2\pi}{\lambda_p}$ normalized by the momentum of free-space light ($q_{air} = \frac{2\pi}{\lambda}$), showing good agreement with the calculated Im $r_p$ (red curve) based on experimental dielectric constants of ZrSiSe.

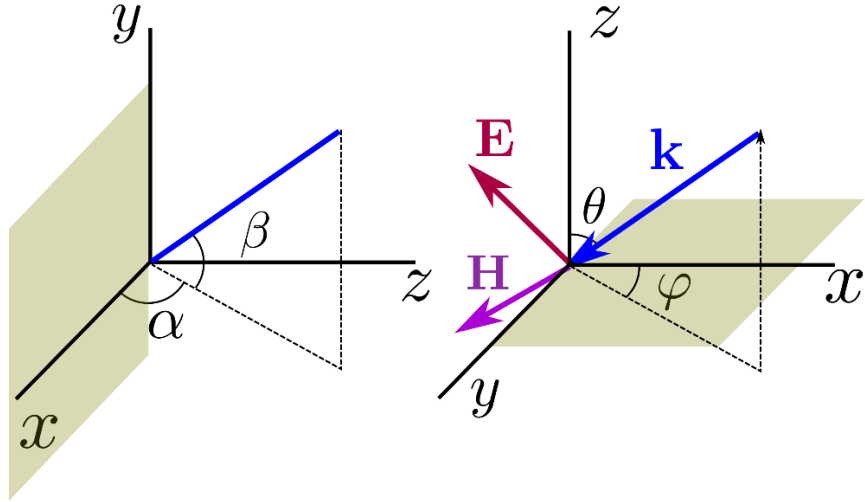

**Fig. S16. Simulation coordinates.** A schematic illustrating different choices of coordinate systems used in the expressions for the scattered fields. The shaded region represents the conducting screen creating the diffraction pattern, with the screen running parallel to the *y*-axis.

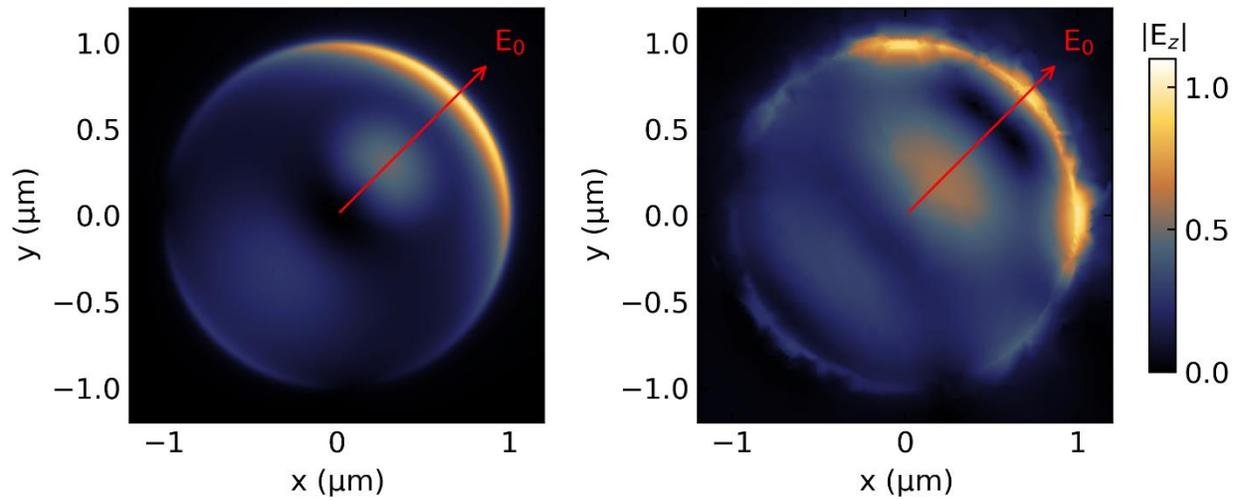

**Fig. S17. Simulation of z-axis electric field of a bare Au disk antenna.** Absolute value of the z-component of the scattered field $E_z$ at a height of 25 nm above the disk, obtained using the approximate model (Left) and the numerical solution (Right). The red arrow in both panels highlights the direction of the incident field.

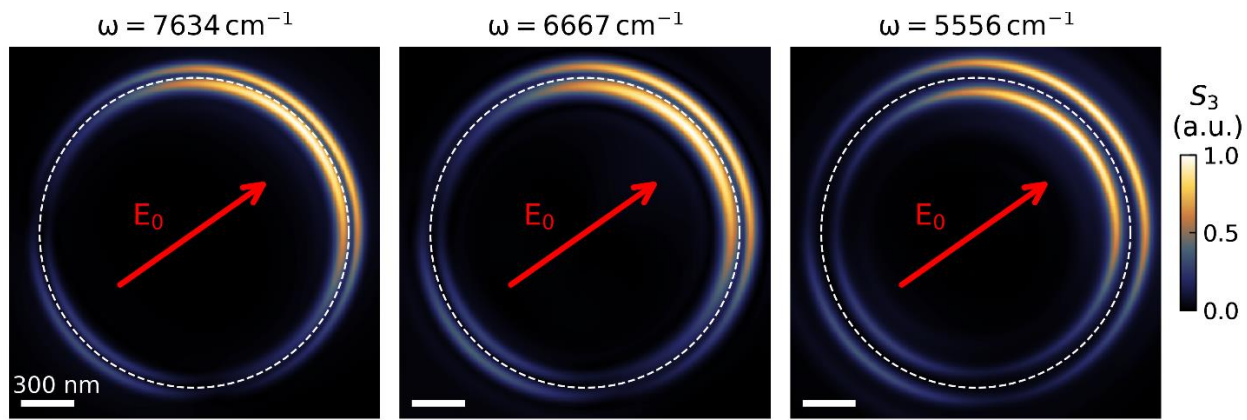

**Fig. S18. Hyperbolic polariton antenna launching simulation.** Simulated near-field amplitude of ZrSiSe (26.5 nm) on a gold disk (25 nm) obtained from the approximate model at $\omega = 7634\ cm^{-1}, 6667\ cm^{-1}, 5556\ cm^{-1}$. The white dashed line shows the edge of the gold disk and the red arrow indicates the direction of the field.

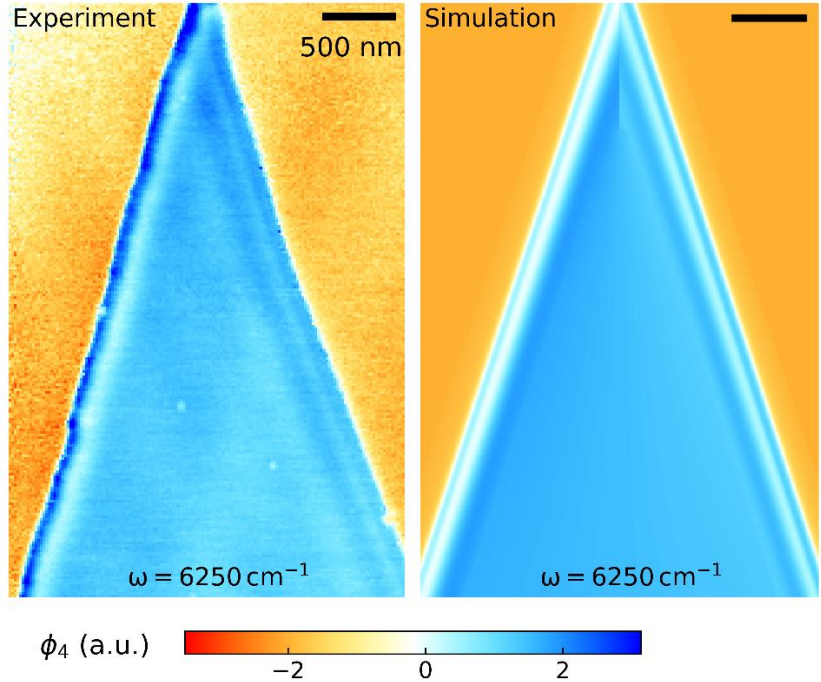

**Fig. S19. Angular dependence in triangular-shaped thin crystal of ZrSiSe.**
(Left) Experimental phase-contrast image ($\phi_4$) of hyperbolic plasmon polaritons near the edges of the crystal at $\omega = 6250$ cm$^{-1}$, showing apparent differences in fringe spacing for two edges of the flake. (Right) Simulation of the phase contrast $\phi_4$ image using the same polartion wavelengths: $\lambda_{p0} = 580\ nm$ and $\lambda_{p1} = 140\ nm$ for the principal ($q_0$) and higher-order ($q_1$) mode, respectively.

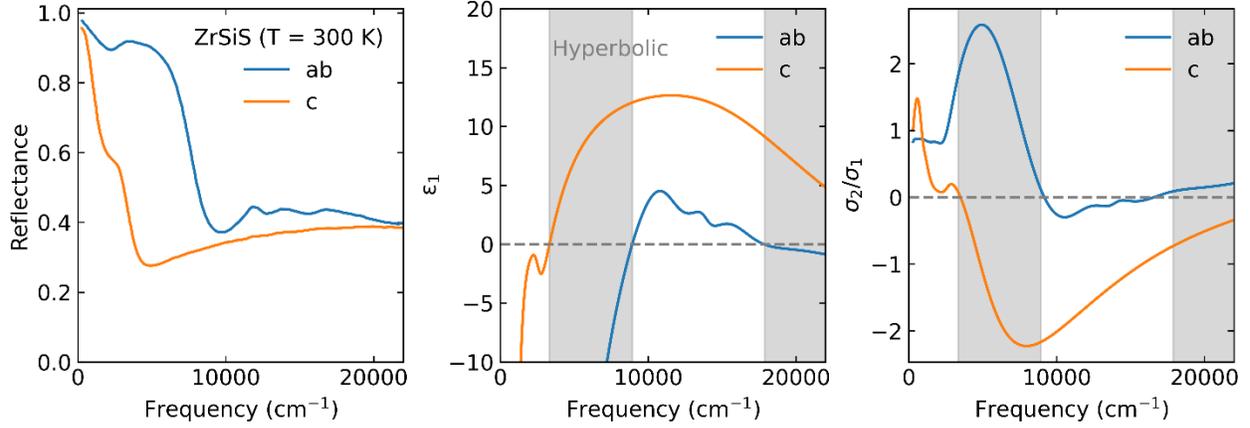

**Fig. S20. ZrSiS.** Anisotropic reflectance, real part of the dielectric function and $\frac{\sigma_2}{\sigma_1}$ ratio of ZrSiS (*7*). Gray shaded regions indicate the hyperbolic frequency regime.

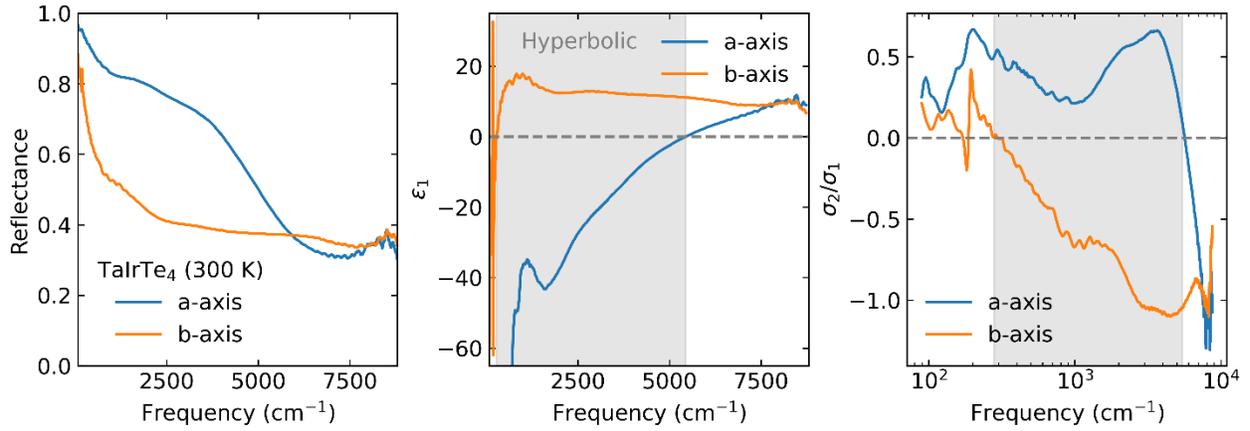

**Fig. S21. TaIrTe$_4$**. Anisotropic reflectance, real part of the dielectric function and $\frac{\sigma_2}{\sigma_1}$ ratio of TaIrTe$_4$ (*15*).

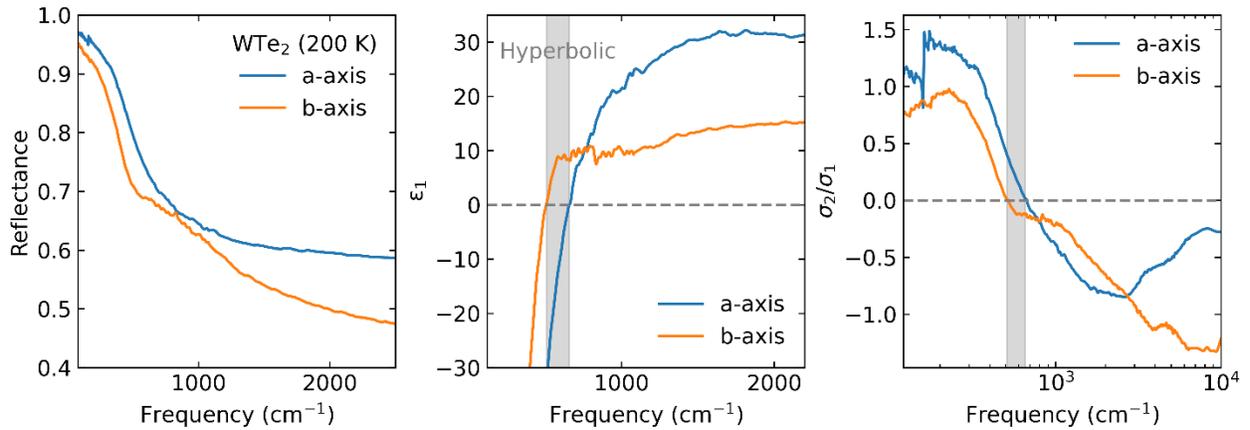

**Fig. S22. WTe$_2$**. Anisotropic reflectance, real part of the dielectric function and $\frac{\sigma_2}{\sigma_1}$ ratio of WTe$_2$ (*16*).

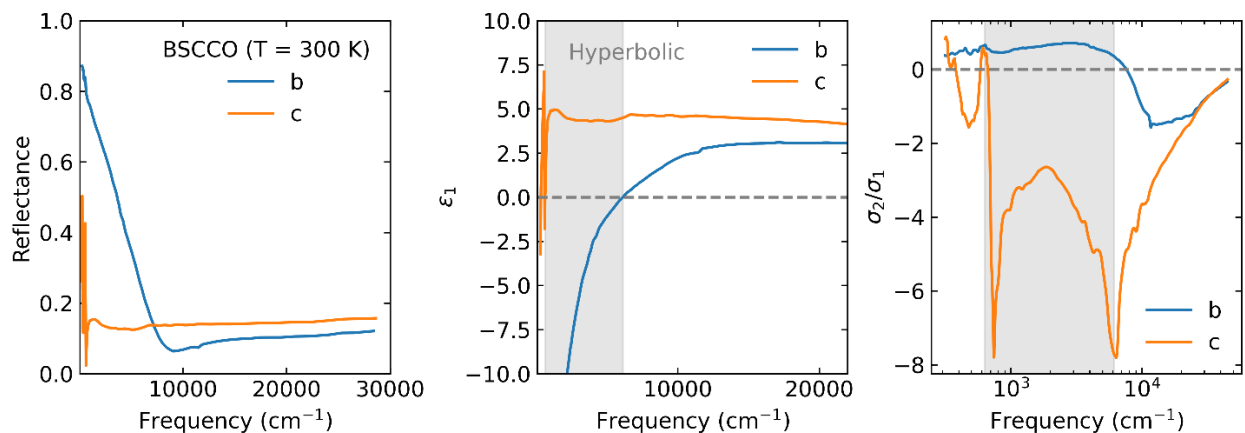

**Fig. S23. $Bi_2Sr_2CaCu_2O_8$**. Anisotropic reflectance, real part of the dielectric function and $\frac{\sigma_2}{\sigma_1}$ ratio of $Bi_2Sr_2CaCu_2O_8$ (BSCCO) (*17*).

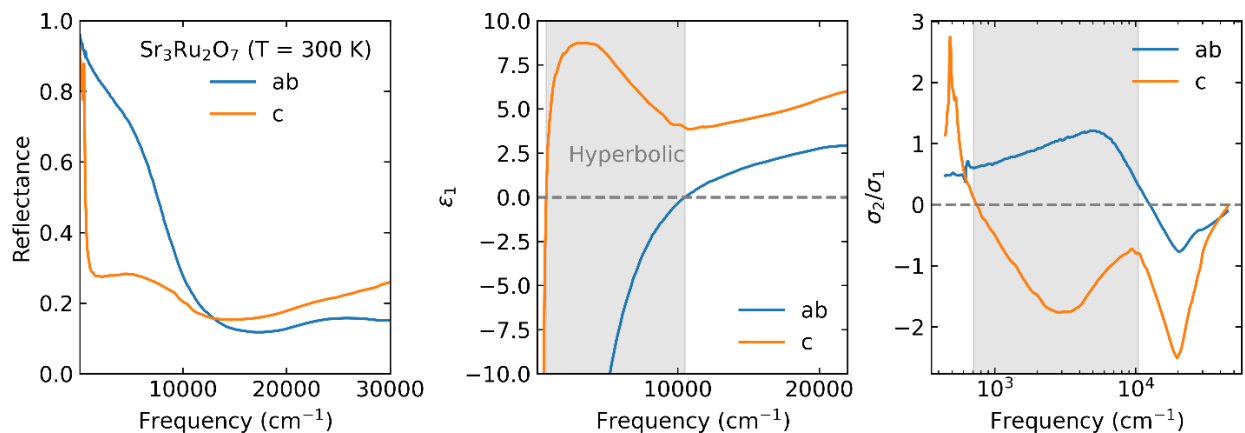

**Fig. S24. $Sr_3Ru_2O_7$.** Anisotropic reflectance, real part of the dielectric function and $\frac{\sigma_2}{\sigma_1}$ ratio of $Sr_3Ru_2O_7$ (*18*).

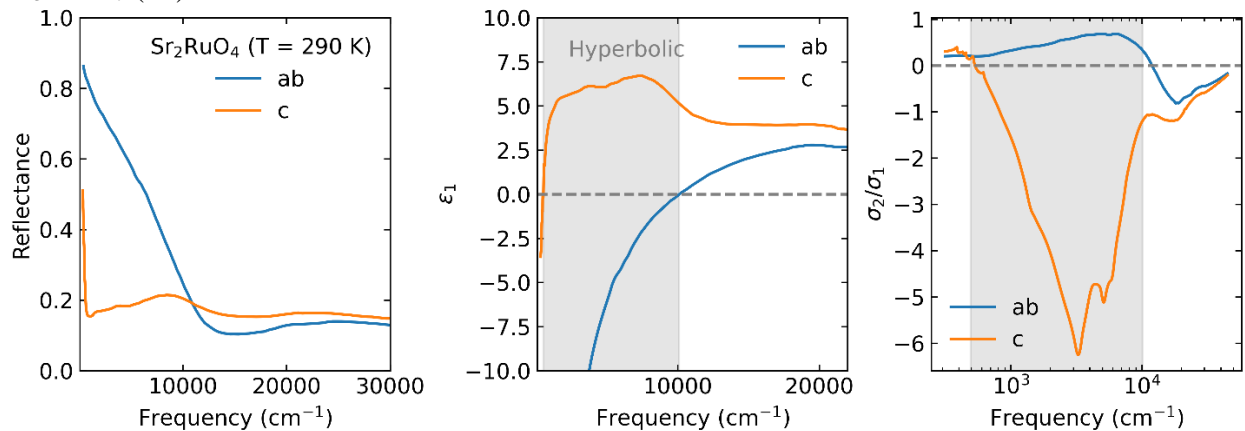

**Fig. S25. $Sr_2RuO_4$.** Anisotropic reflectance, real part of the dielectric function and $\frac{\sigma_2}{\sigma_1}$ ratio of $Sr_2RuO_4$ (*19*).

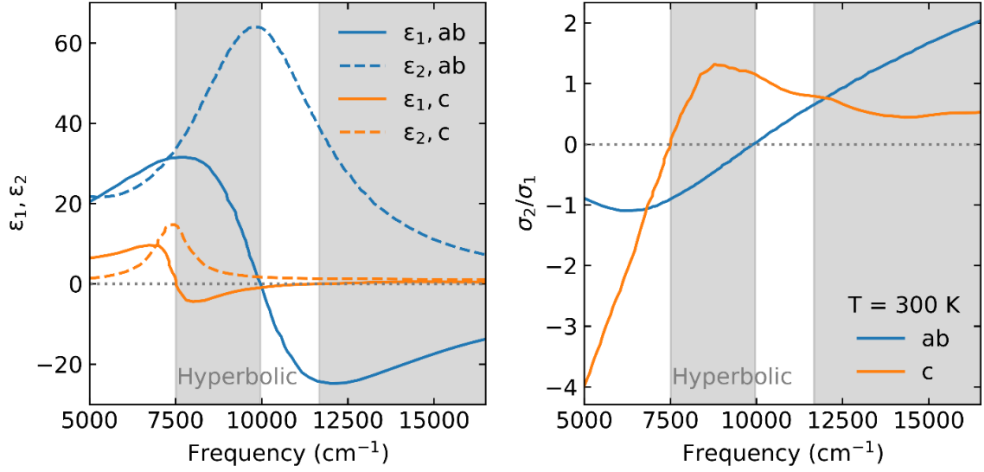

**Fig. S26. Bi$_2$Te$_3$.** Anisotropic dielectric function and $\frac{\sigma_2}{\sigma_1}$ ratio of Bi$_2$Te$_3$ (*20*).

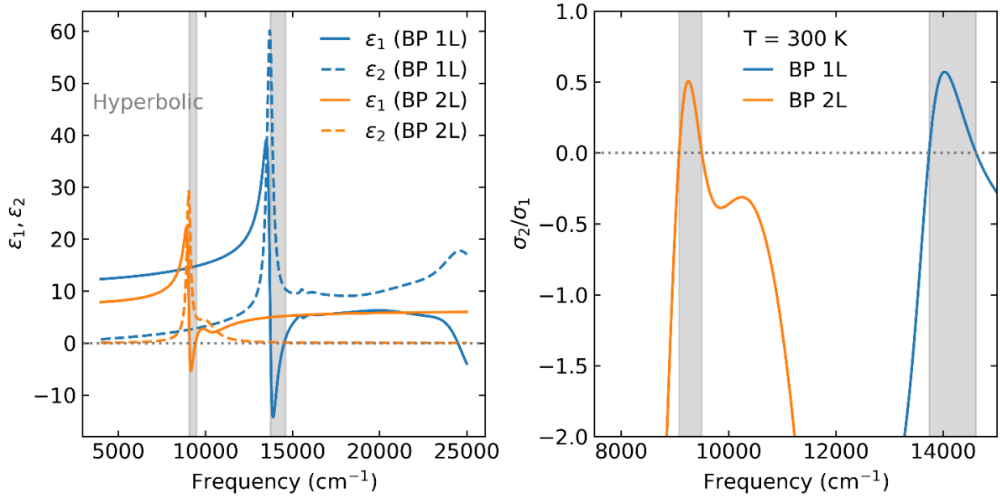

**Fig. S27. Black phosphorus.** Excitonic dielectric function of monolayer (1L) and bilayer (2L) black phosphorus along the arm-chair direction (*21*) and the corresponding $\frac{\sigma_2}{\sigma_1}$ ratio.

| j | $\omega_{0,j}$ $(cm^{-1})$ | $\omega_{p,j}$ $(cm^{-1})$ | $\gamma_j$ $(cm^{-1})$ |
|---|---|---|---|
| 1 | 0 | 5127.2 | 400.0 |
| 2 | 6291.3 | 2057.4 | 715.2 |
| 3 | 7367.8 | 2242.8 | 1055.3 |

**Table S1**. Parameters for the Drude-Lorentz model fitting of the experimental out-of-plane dielectric function of ZrSiSe using $\varepsilon_c(\omega) = \varepsilon_\infty + \sum_j \omega_{p,j}^2/(\omega_{0,j}^2 - \omega^2 - i\gamma_j\omega)$. Here, $\varepsilon_\infty$=2.96 is the high frequency dielectric constant.

**Reference**

1. J. Hu, Z. Tang, J. Liu, X. Liu, Y. Zhu, D. Graf, K. Myhro, S. Tran, C. N. Lau, J. Wei, Z. Mao, Evidence of Topological Nodal-Line Fermions in ZrSiSe and ZrSiTe. *Phys. Rev. Lett.* **117**, 016602 (2016).

2. Y. Shao, A. N. Rudenko, J. Hu, Z. Sun, Y. Zhu, S. Moon, A. J. Millis, S. Yuan, A. I. Lichtenstein, D. Smirnov, Z. Q. Mao, M. I. Katsnelson, D. N. Basov, Electronic correlations in nodal-line semimetals. *Nat. Phys.* **16**, 636–641 (2020).

3. N. Tancogne-Dejean, A. Rubio, Parameter-free hybridlike functional based on an extended Hubbard model: $\mathrm{DFT}+U+V$. *Phys. Rev. B*. **102**, 155117 (2020).

4. N. Tancogne-Dejean, M. J. T. Oliveira, X. Andrade, H. Appel, C. H. Borca, G. Le Breton, F. Buchholz, A. Castro, S. Corni, A. A. Correa, U. De Giovannini, A. Delgado, F. G. Eich, J. Flick, G. Gil, A. Gomez, N. Helbig, H. Hübener, R. Jestädt, J. Jornet-Somoza, A. H. Larsen, I. V. Lebedeva, M. Lüders, M. A. L. Marques, S. T. Ohlmann, S. Pipolo, M. Rampp, C. A. Rozzi, D. A. Strubbe, S. A. Sato, C. Schäfer, I. Theophilou, A. Welden, A. Rubio, Octopus, a computational framework for exploring light-driven phenomena and quantum dynamics in extended and finite systems. *J. Chem. Phys.* **152**, 124119 (2020).

5. G. Gatti, A. Crepaldi, M. Puppin, N. Tancogne-Dejean, L. Xian, U. De Giovannini, S. Roth, S. Polishchuk, Ph. Bugnon, A. Magrez, H. Berger, F. Frassetto, L. Poletto, L. Moreschini, S. Moser, A. Bostwick, E. Rotenberg, A. Rubio, M. Chergui, M. Grioni, Light-Induced Renormalization of the Dirac Quasiparticles in the Nodal-Line Semimetal ZrSiSe. *Phys. Rev. Lett.* **125**, 076401 (2020).

6. C. Hartwigsen, S. Goedecker, J. Hutter, Relativistic separable dual-space Gaussian pseudopotentials from H to Rn. *Phys. Rev. B*. **58**, 3641–3662 (1998).

7. J. Ebad-Allah, S. Rojewski, M. Vöst, G. Eickerling, W. Scherer, E. Uykur, R. Sankar, L. Varrassi, C. Franchini, K.-H. Ahn, J. Kuneš, C. A. Kuntscher, Pressure-Induced Excitations in the Out-of-Plane Optical Response of the Nodal-Line Semimetal ZrSiS. *Phys. Rev. Lett.* **127**, 076402 (2021).

8. B. Xu, Y. M. Dai, L. X. Zhao, K. Wang, R. Yang, W. Zhang, J. Y. Liu, H. Xiao, G. F. Chen, A. J. Taylor, D. A. Yarotski, R. P. Prasankumar, X. G. Qiu, Optical spectroscopy of the Weyl semimetal TaAs. *Phys. Rev. B*. **93**, 121110 (2016).

9. C. J. Tabert, J. P. Carbotte, Optical conductivity of Weyl semimetals and signatures of the gapped semimetal phase transition. *Phys. Rev. B*. **93**, 085442 (2016).

10. M. B. Schilling, L. M. Schoop, B. V. Lotsch, M. Dressel, A. V. Pronin, Flat Optical Conductivity in ZrSiS due to Two-Dimensional Dirac Bands. *Phys. Rev. Lett.* **119**, 187401 (2017).

11. T. Habe, M. Koshino, Dynamical conductivity in the topological nodal-line semimetal ZrSiS. *Phys. Rev. B*. **98**, 125201 (2018).

12. A. Sommerfeld, O. Laporte, P. A. Moldauer, Optics: Vol. 5 of Lectures on Theoretical Physics. *Physics Today*. **8**, 16–16 (1955).

13. P. C. Clemmow, D. R. Hartree, A method for the exact solution of a class of two-dimensional diffraction problems. *Proceedings of the Royal Society of London. Series A. Mathematical and Physical Sciences*. **205**, 286–308 (1951).

14. S. Dai, Q. Ma, T. Andersen, A. S. Mcleod, Z. Fei, M. K. Liu, M. Wagner, K. Watanabe, T. Taniguchi, M. Thiemens, F. Keilmann, P. Jarillo-Herrero, M. M. Fogler, D. N. Basov, Subdiffractional focusing and guiding of polaritonic rays in a natural hyperbolic material. *Nat. Commun.* **6**, 6963 (2015).

15. Y. Shao, R. Jing, S. H. Chae, C. Wang, Z. Sun, E. Emmanouilidou, S. Xu, D. Halbertal, B. Li, A. Rajendran, F. L. Ruta, L. Xiong, Y. Dong, A. S. McLeod, S. S. Sunku, J. C. Hone, J. Moore, J. Orenstein, J. G. Analytis, A. J. Millis, N. Ni, D. Xiao, D. N. Basov, Nonlinear nanoelectrodynamics of a Weyl metal. *Proc. Natl. Acad. Sci. U. S. A.* **118**, e2116366118 (2021).

16. A. J. Frenzel, C. C. Homes, Q. D. Gibson, Y. M. Shao, K. W. Post, A. Charnukha, R. J. Cava, D. N. Basov, Anisotropic electrodynamics of type-II Weyl semimetal candidate ${\mathrm{WTe}}_{2}$. *Phys. Rev. B*. **95**, 245140 (2017).

17. S. Tajima, G. D. Gu, S. Miyamoto, A. Odagawa, N. Koshizuka, Optical evidence for strong anisotropy in the normal and superconducting states in ${\mathrm{Bi}}_{2}${\mathrm{Sr}}_{2}${\mathrm{CaCu}}_{2}${\mathrm{O}}_{8+\mathit{z}}$. *Phys. Rev. B*. **48**, 16164–16167 (1993).

18. C. Mirri, L. Baldassarre, S. Lupi, M. Ortolani, R. Fittipaldi, A. Vecchione, P. Calvani, Anisotropic optical conductivity of ${\text{Sr}}_{3}{\text{Ru}}_{2}{\text{O}}_{7}$. *Phys. Rev. B*. **78**, 155132 (2008).

19. T. Katsufuji, M. Kasai, Y. Tokura, In-Plane and Out-of-Plane Optical Spectra of ${\mathrm{Sr}}_{2}{\mathrm{RuO}}_{4}$. *Phys. Rev. Lett.* **76**, 126–129 (1996).

20. M. Esslinger, R. Vogelgesang, N. Talebi, W. Khunsin, P. Gehring, S. de Zuani, B. Gompf, K. Kern, Tetradymites as Natural Hyperbolic Materials for the Near-Infrared to Visible. *ACS Photonics*. **1**, 1285–1289 (2014).

21. F. Wang, C. Wang, A. Chaves, C. Song, G. Zhang, S. Huang, Y. Lei, Q. Xing, L. Mu, Y. Xie, H. Yan, Prediction of hyperbolic exciton-polaritons in monolayer black phosphorus. *Nat. Comm.* **12**, 5628 (2021).